\documentclass[english,aps,twocolumn,prl,superscriptaddress]{revtex4-1}
\usepackage[T1]{fontenc}
\usepackage[latin9]{inputenc}
\usepackage{color}
\usepackage{babel}
\usepackage{booktabs}
\usepackage{amssymb}
\usepackage{graphicx}
\usepackage{esint}
\usepackage[unicode=true,pdfusetitle,
 bookmarks=true,bookmarksnumbered=false,bookmarksopen=false,
 breaklinks=false,pdfborder={0 0 1},backref=false,colorlinks=false]
 {hyperref}

\makeatletter

\providecommand{\tabularnewline}{\\}

\makeatother

\begin{document}

\title{Cavity Optomechanics in a Levitated Helium Drop}

\author{L.~Childress}

\affiliation{Department of Physics, Yale University, New Haven, CT, 06520, USA}

\affiliation{Department of Physics, McGill University, 3600 Rue University, Montreal,
Quebec H3A 2T8, Canada}

\author{M.~P.~Schmidt}

\affiliation{Institute for Theoretical Physics, Department of Physics, University
of Erlangen-Nürnberg, Staudtstrasse 7, 91058 Erlangen, Germany}

\author{A.~D.~Kashkanova}

\affiliation{Department of Physics, Yale University, New Haven, CT, 06520, USA}

\author{C.~D.~Brown}

\affiliation{Department of Physics, Yale University, New Haven, CT, 06520, USA}

\author{G.~I.~Harris}

\affiliation{Department of Physics, Yale University, New Haven, CT, 06520, USA}

\author{A.~Aiello}

\affiliation{Institute for Theoretical Physics, Department of Physics, University
of Erlangen-Nürnberg, Staudtstrasse 7, 91058 Erlangen, Germany}

\affiliation{Max Planck Institute for the Science of Light, Staudtstr. 2, 91058
Erlangen, Germany}

\author{F. Marquardt}

\affiliation{Institute for Theoretical Physics, Department of Physics, University
of Erlangen-Nürnberg, Staudtstrasse 7, 91058 Erlangen, Germany}

\affiliation{Max Planck Institute for the Science of Light, Staudtstr. 2, 91058
Erlangen, Germany}

\author{J. G. E. Harris}

\affiliation{Department of Physics, Yale University, New Haven, CT, 06520, USA}

\affiliation{Department of Applied Physics, Yale University, New Haven, CT, 06520,
USA}

\affiliation{Yale Quantum Institute, Yale University, New Haven, CT, 06520, USA}
\begin{abstract}
We describe a proposal for a new type of optomechanical system based
on a drop of liquid helium that is magnetically levitated in vacuum.
In the proposed device, the drop would serve three roles: its optical
whispering gallery modes would provide the optical cavity, its surface
vibrations would constitute the mechanical element, and evaporation
of ${\rm He}$ atoms from its surface would provide continuous refrigeration.
We analyze the feasibility of such a system in light of previous experimental
demonstrations of its essential components: magnetic levitation of
mm-scale and cm-scale drops of liquid ${\rm He}$, evaporative cooling
of He droplets in vacuum, and coupling to high-quality optical whispering
gallery modes in a wide range of liquids. We find that the combination
of these features could result in a device that approaches the single-photon
strong coupling regime, due to the high optical quality factors attainable
at low temperatures. Moreover, the system offers a unique opportunity
to use optical techniques to study the motion of a superfluid that
is freely levitating in vacuum (in the case of $^{4}{\rm He}$). Alternatively,
for a normal fluid drop of $^{{\rm 3}}{\rm He}$, we propose to exploit
the coupling between the drop's rotations and vibrations to perform
quantum non-demolition measurements of angular momentum.
\end{abstract}
\maketitle

\subsection*{Introduction}

Optomechanical systems \cite{aspelmeyer_cavity_2014} have been used
to demonstrate quantum effects in the harmonic motion of macroscopic
objects over a very broad range of physical regimes. For example,
quantum optomechanical effects have been observed in the motion of
objects formed from all three states of matter (solid \cite{oconnell_quantum_2010},
gas \cite{brahms_optically_2012}, and liquid \cite{kashkanova_quantum_2017});
at temperatures ranging from cryogenic to room temperature \cite{purdy_quantum_2017};
with effective mass as large as $\sim$100 nanograms \cite{underwood_measurement_2015};
and with resonance frequencies ranging from kHz to GHz. Despite rapid
progress, a number of important goals in this field remain outstanding,
for example generating highly non-classical states\textcolor{red}{{}
}of motion with negative quasiprobability distributions or which violate
a Bell-type inequality (even without postselection); efficiently transferring
quantum states between microwave and optical frequencies; and observing
quantum effects in the motion of objects massive enough to constrain
theories of quantum gravity \cite{pikovski_probing_2012}. Access
to these phenomena may be facilitated by devices with reduced optical
and mechanical loss, increased optomechanical coupling, and increased
mass. In addition, new regimes and qualitatively new forms of optomechanical
coupling may be accessed by developing systems in which the mechanical
degrees of freedom are not simply the harmonic oscillations of an
elastic body. In this work, we will show that a levitated drop of
superfluid helium will be a most promising platform that combines
many of these desired features and offers novel possibilities.

To date, most optomechanical devices are realized by using solid objects
(e.g., mirrors, waveguides, or electrical circuits) to confine modes
of the electromagnetic field, and ensuring that these confined modes
couple to the harmonic motion of a solid object. These devices' performance
is determined in part by the properties of the solids from which they
are formed. For example, the material\textquoteright s mechanical
and electromagnetic loss are important parameters, as is the material\textquoteright s
compatibility with fabrication techniques. For this reason, high-quality
dielectrics are typically employed in devices using optical fields,
while superconductors are typically employed in devices using microwave
fields. 

\begin{figure}
\includegraphics[width=1\columnwidth]{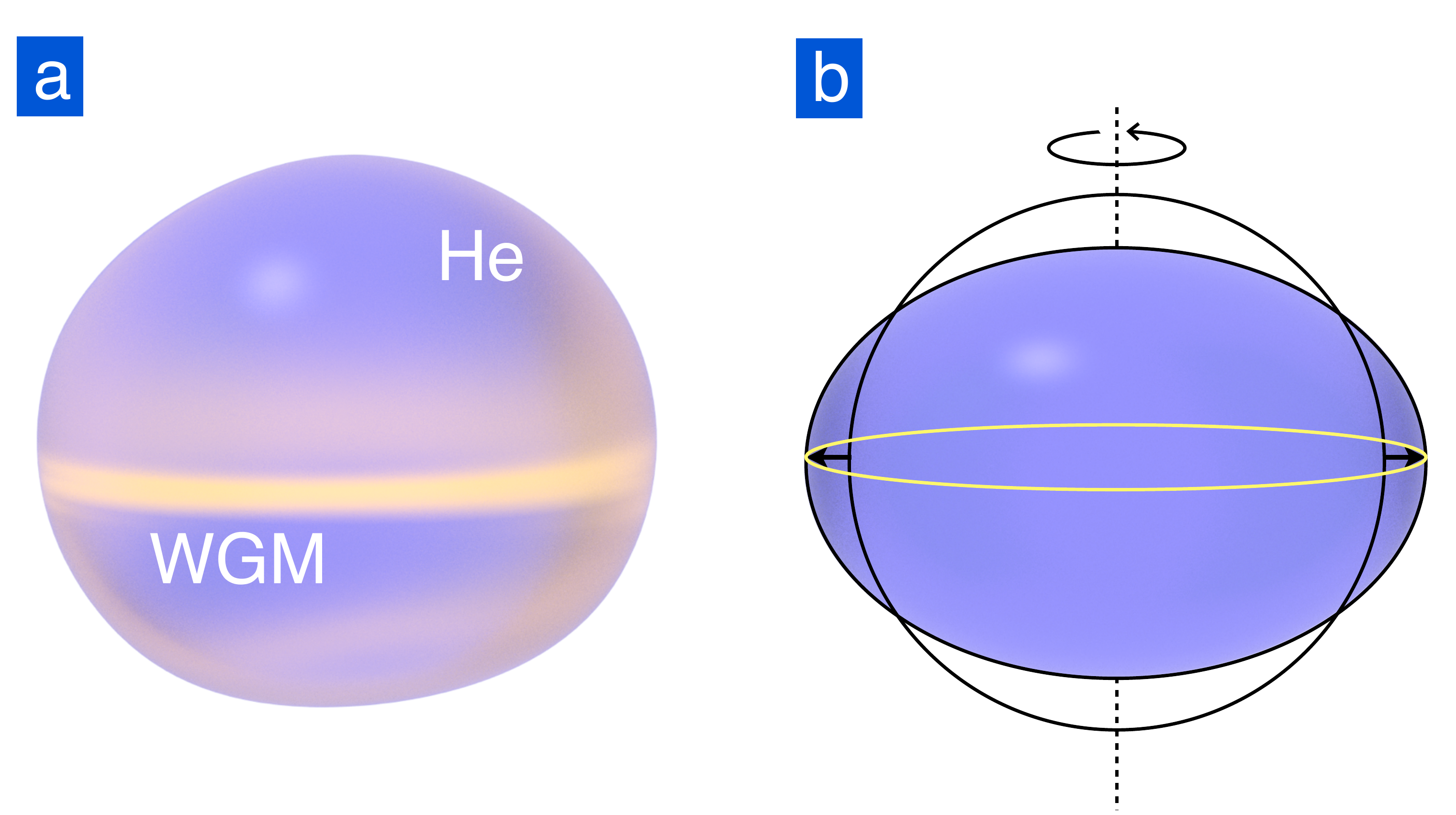}

\caption{\label{SetupFigure}(a) Schematic illustration of a levitated helium
drop containing an optical whispering gallery mode (WGM), whose optical
path length is modified by the surface vibrations. (b) Rotation of
the drop leads to an equatorial bulge, which also modifies the WGM's
path length.}
\end{figure}

Most solid-based optomechanical devices must be placed in direct contact
with their solid surroundings, both to support them against Earth\textquoteright s
gravity and to provide thermal anchoring. This contact can negatively
affect the device\textquoteright s performance, as it represents a
route for mechanical loss. It may also be problematic if the contact
is not able to provide effective cooling (i.e., to counteract heating
from electromagnetic absorption in the device), as elevated temperatures
tend to obscure quantum effects. 

If the mechanical element is a solid object that is levitated in vacuum
(e.g., using optical or magnetic forces) \cite{chang_cavity_2010,romero-isart_toward_2010,barker_cavity_2010,millen_nanoscale_2014,li_measurement_2010,li_millikelvin_2011,romero-isart_large_2011,gieseler_subkelvin_2012},
the absence of direct contact can result in very low loss for some
mechanical degrees of freedom (particularly the object\textquoteright s
center-of-mass). However the absence of direct contact also precludes
effective cooling of the element. This is particularly important given
the non-zero optical absorption of conventional materials and the
high optical powers typically required for levitation and/or read
out of the object\textquoteright s motion. As a result, solid objects
levitated in vacuum have operated at elevated bulk temperatures (although
some degrees of freedom may still be cooled to very low effective
temperatures). 

In contrast to solid objects, atomic gases may be levitated and trapped
in vacuum at very low temperatures. This is due to two important features
of atomic systems: first, the gas is heated only via the atoms\textquoteright{}
spontaneous emission (which can be minimized by using laser fields
that are far detuned from the atomic transitions). Second, the atoms
can all be kept cold by laser cooling and evaporation. When a cloud
of ultracold atoms is trapped inside an optical cavity, its center-of-mass
motion (or some collective mode of the gas) can detune the cavity,
leading to an optomechanical interaction \cite{murch_observation_2008,brennecke_cavity_2008}.
This interaction may be quite strong, as the small number of atoms
can be compensated by the cloud's large zero point motion and by adjusting
the detuning between the atomic transition and the cavity. Ultracold
atom-based optomechanical devices have achieved optomechanical figures
of merit and demonstrated quantum optomechanical effects that are
competitive with state-of-the-art solid-based devices. However the
effective mass of atom-based devices is likely to remain several orders
of magnitude lower than solid-based devices, making them less promising
for foundational tests.

Recently, optomechanical devices that employ liquids have been demonstrated.
These can be realized by supporting a drop of liquid \cite{dahan_droplet_2016}
so that its free surface confines an electromagnetic mode in the form
of an optical whispering gallery mode (WGM). In this case, the drop
serves as both the optical cavity and the mechanical element, as the
drop\textquoteright s surface oscillations tend to detune the drop\textquoteright s
optical WGMs. Devices based on this approach have been demonstrated
at room temperature and with the drops mechanically anchored (rather
than levitating). However, the relatively high mechanical loss in
room-temperature fluids has precluded them from accessing quantum
optomechanical effects. 

Liquid-based optomechanical devices can also be realized by filling
\cite{bahl_brillouin_2013,kashkanova_superfluid_2017,kashkanova_optomechanics_2017,lorenzo_superfluid_2014}
or coating \cite{harris_laser_2016} a solid electromagnetic cavity
with a fluid. In this case only the mechanical degree of freedom is
provided by the fluid, for example as a density wave or surface wave
that detunes the cavity by modulating the overlap between the liquid
and the cavity mode. This approach has been used at cryogenic temperatures
with superfluid ${\rm ^{4}He}$ serving as the liquid \cite{lorenzo_superfluid_2014,harris_laser_2016,kashkanova_superfluid_2017,kashkanova_optomechanics_2017,singh_detecting_2016}.

Liquid He has a number of properties that make it appealing for optomechanical
devices. Its large bandgap ($\sim19{\rm \:eV}$), chemical purity,
and lack of structural defects should provide exceptionally low electromagnetic
loss. In its pure superfluid state, the viscosity that strongly damps
other liquids is absent. The mechanical loss arising from its nonlinear
compressibility varies with temperature $T$ as $T^{4}$, and so is
strongly suppressed at low $T$. In addition, its thermal conductivity
at cryogenic temperatures is exceptionally large. 

To date, optomechanical devices based on superfluid-filled cavities
have reaped some advantage from these features (including the observation
of quantum optomechanical effects \cite{kashkanova_quantum_2017}).
However the need to confine the superfluid within a solid vessel has
undercut many of the advantages offered by superfluid helium. This
is because direct contact between the superfluid and a solid object
provides a channel for mechanical losses (i.e., radiation of mechanical
energy from the superfluid into the solid) and heating (due to electromagnetic
absorption in the solid). 

In this paper, we propose a new type of optomechanical device that
is intended to combine advantages from each type of device described
above. Specifically, we consider a millimeter-scale drop of superfluid
He that is magnetically levitated in vacuum (Fig.~\ref{SetupFigure}).
Magnetic levitation would provide high-quality optical WGMs and high-quality
mechanical modes by confining the optical and mechanical energy entirely
within the superfluid. Despite being levitated in vacuum, the drop
would be able to cool itself efficiently by evaporation, thereby compensating
for any residual heating. 

In addition to offering these technical improvements, this approach
would provide access to qualitatively new forms of optomechanical
coupling. A levitated drop of $^{3}{\rm He}$ in its normal state
would retain the low optical loss and efficient cooling of the superfluid
drop, but would experience viscous damping of its normal modes of
oscillation. However its rigid body rotation (which is not directly
damped by viscosity) would couple to the drop\textquoteright s optical
WGMs. The coupling arising in such an \textquotedblleft opto-rotational\textquotedblright{}
system is distinct from the usual optomechanical coupling, with important
consequences for quantum effects. 

Besides establishing a novel optomechanics platform, the proposed
system may also help address long-standing questions regarding the
physics of liquid helium. For example, a levitated drop of $^{4}{\rm He}$
may contain a vortex line \cite{gomez_shapes_2014,bauer_vortex_1995}
which deforms the drop shape and hence detunes the optical WGMs, providing
a probe of vortex dynamics. Alternately, optical measurements of a
levitated drop could probe the onset and decay of turbulence in a
system without walls. 

Most of the essential features of the proposed device have been demonstrated
previously, albeit in disparate settings. These include: the magnetic
levitation and trapping of mm-scale and cm-scale drops of superfluid
helium \cite{weilert_magnetic_1996,weilert_magnetic_1997}, the characterization
of these drops\textquoteright{} surface modes \cite{vicente_surface_2002}
for $T>650\:{\rm mK}$; the observation of evaporative cooling of
He drops \cite{toennies_superfluid_2004} in vacuum, and the observation
of high-finesse optical WGMs in liquids such as ethanol \cite{tzeng_laser_1984,tzeng_laser-induced_1985,qian_lasing_1986}
and water \cite{tanyeri_lasing_2007} (at room temperature) and in
liquid $H_{2}$ \cite{uetake_stimulated_1999,uetake_stimulated_2002}
(at $T\sim15$ K). This paper uses these prior results to estimate
the optomechanical properties of a levitated drop of liquid He, including
the possible coupling to rotational motion. The discussion presented
here is relevant for both ${\rm ^{3}He}$ and ${\rm ^{4}He}$, except
where noted otherwise. 

\section{Optomechanical Coupling in a Helium Drop}

We begin by discussing the vibrational modes of the drop and deriving
their optomechanical coupling to the optical WGMs. Note that WGMs
in spherical (and near-spherical) dielectrics are discussed extensively
in the literature \cite{oraevsky_whispering-gallery_2002}, so we
do not review their properties here. 

\subsection{Vibrational modes}

\begin{figure}
\includegraphics[width=1\columnwidth]{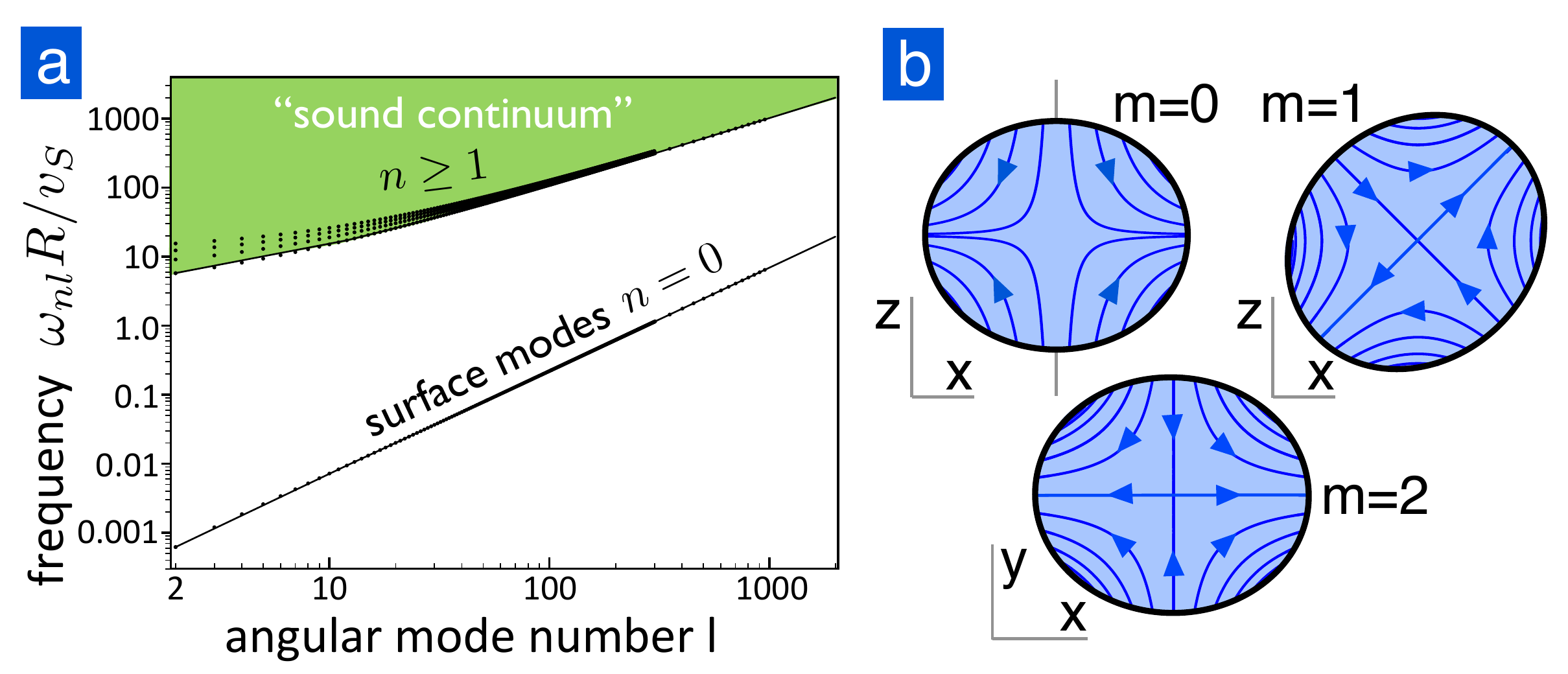}

\caption{\label{FigureVibrationalModes}(a) Vibrational modes of a spherical
drop, with radial index $n$ and angular index $l$; the surface modes
$n=0$ are separated from the bulk continuum ($n\geq1$). \textcolor{black}{Points
represent the discrete mode frequencies, and solid lines represent
the analytical expressions for the $n=0$ and $n=1$ mode frequencies.}\textcolor{red}{{}
}(b) Illustrations of the velocity profiles for $l$ = 2 surface modes
of different azimuthal number $m$.}
\end{figure}

The vibrational modes of a helium drop can be calculated by solving
the linearized hydrodynamic equations (Fig.~\ref{FigureVibrationalModes}).
The angular dependence of each mode is given by a spherical harmonic
$Y_{l,m}(\theta,\phi)$ (where $l$ and $m$ index the mode's total
angular momentum and its projection on the $z$-axis). The radial
dependence of each mode can be written in terms of spherical Bessel
functions $j_{n}(kr)$ (where $k$ is the mode's wavenumber and $n$
determines the number of radial nodes). The physical nature of these
modes falls into two classes:

(i) Low-frequency surface modes (ripplons), whose restoring force
is provided by surface tension. These have frequency $\omega_{l}=\sqrt{l(l-1)(l+2)\sigma/(\rho R^{3})}$
\cite{lord_rayleigh_capillary_1879} for the $2l+1$ degenerate modes
at any given angular mode number $l=2,3,\ldots$ , where $R$ is the
radius of the drop, $\rho$ is its density, and $\sigma$ is its surface
tension. For a $^{4}{\rm He}$ drop of radius $R=1\:{\rm mm}$, the
$l=2$ mode whose optomechanical coupling we will analyze has a frequency
of $\omega_{2}=2\pi\cdot23\,$ Hz $\equiv\omega_{\mathrm{{vib}}}$.

(ii) Sound modes, whose restoring force is provided by the elastic
modulus. The frequency of these modes depends on the indices $n$
and $l$ \cite{bohr_nuclear_1998,schmidt_optomechanical_2015}. These
include ``breathing'' modes and acoustic whispering gallery modes,
among others. Their frequencies scale with $v_{s}/R$ where $v_{s}$
is the speed of sound in liquid He. For the example of a $^{4}{\rm He}$
drop with $R=1\:{\rm mm}$, the lowest-frequency compressional mode
oscillates at $2\pi\cdot120$ kHz. 

In the present work we focus on the surface modes, specifically the
lowest nontrivial modes (quadrupole deformations, $l=2$). These couple
most strongly to the optical WGMs.

\subsection{Optomechanical Coupling to Surface Modes}

The single-quantum optomechanical coupling can be found from the optical
WGM detuning produced by the surface mode's quantum zero-point fluctuation
amplitude. To calculate this amplitude, we note that the surface deflection
$\delta R(\theta,\varphi)$ can be decomposed in terms of the surface
modes as $\delta R=\sum_{l,m}X_{l,m}Y_{l,m}(\theta,\varphi)$, where
$X_{l,-m}=X_{l,m}^{*}$ are the time-dependent mode amplitudes. The
spherical harmonics $Y_{l,m}$ are normalized such that $\int d\Omega\,\left|Y_{l,m}\right|^{2}=1$.

The potential energy of the modes is determined by surface tension
$\sigma$. For the $l=2$ modes of interest here, the increase of
surface area is given (to lowest order) by $2\sum_{m}\left|X_{m}\right|^{2}$.
We note that in order to obtain this result, care needs to be taken
to preserve the volume of the drop by adjusting the radius (i.e. the
$l=0$ monopole contribution to $\delta R$) \cite{gang_thermal_1995}.
Focusing on the $l=2,m=0$ mode, we then equate the average potential
energy $2\sigma\left\langle X_{0}^{2}\right\rangle $ to half of the
zero-point energy $\hbar\omega_{{\rm vib}}/4$. From this, we find
the zero-point fluctuation amplitude of the $m=0$ surface mode, as
well as the change of radius at the drop's equator:

\begin{equation}
X_{0,{\rm ZPF}}=\sqrt{\frac{\hbar\omega_{{\rm vib}}}{8\sigma}},\:\:\:\:\:\delta R_{{\rm ZPF}}=\sqrt{\frac{5}{16\pi}}X_{0,{\rm ZPF}.}
\end{equation}
Again, for a drop of $^{4}{\rm He}$ with $R=1\,$ mm, this is $X_{0,{\rm ZPF}}=2.2{\rm \:fm}$. 

\begin{figure}
\begin{centering}
\includegraphics[width=0.8\columnwidth]{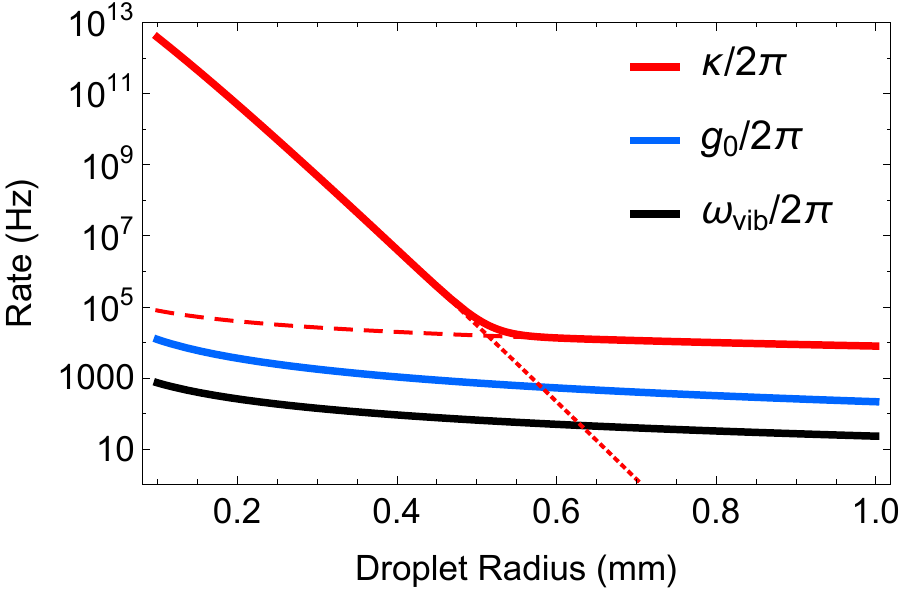}
\par\end{centering}
\caption{\label{Fig3}The mechanical frequency $\omega_{\mathrm{{vib}}}$ for
a $l$=2 mode, the optical decay rate $\kappa$, and the optomechanical
coupling constant $g_{0}$, all as a function of drop radius (for
$T=300\ {\rm mK}$ and $\lambda=1\ \mu{\rm m}$). The dashed curve
shows optical loss due to scattering from thermal surface fluctuations;
the dotted curve shows radiative loss due to surface curvature.}
\end{figure}

Each optical WGM in the drop is specified by the indices $\tilde{l}$,
$\tilde{m}$, and $\tilde{n}$ (which specify the WGM's total angular
momentum, its projection along the $z$-axis, and the number of radial
nodes, respectively). The WGM that lies closest to the drop's equator
(i.e., with $\tilde{l}=\tilde{m}$) has an optical path length that
is proportional to the drop's equatorial circumference. As a consequence,
we find $g_{0}=\omega_{{\rm opt}}\delta R_{{\rm ZPF}}/R$ for the
bare optomechanical coupling between an equatorial optical whispering
gallery mode and the $l=2,\ m=0$ surface mode. For $\lambda=1\ \mu{\rm m}$
and $R=1{\rm \ mm}$, this amounts to $g_{0}=2\pi\cdot213\,{\rm Hz}$
(see Fig. \ref{Fig3}). 

We note that the optical frequency of the equatorial WGM couples linearly
only to the surface mode with $m=0$. All $m\neq0$ vibrational surface
modes will be restricted to (considerably weaker) higher-order coupling.

Optical WGMs with arbitrary ($\tilde{l},\,\tilde{m}$) \cite{lai_time-independent_1990}
also couple linearly to the $l=2,m=0$ mechanical mode, with coupling
rates 

\begin{equation}
g_{0}^{(\tilde{l},\tilde{m})}=\omega_{{\rm opt}}\frac{\delta R_{{\rm ZPF}}}{R}\frac{1}{2}\left[3\frac{\tilde{m}^{2}}{\tilde{l}(\tilde{l}+1)}-1\right].\label{eq:GeneralCouplingConstant}
\end{equation}
WGMs propagating near the equator (i.e., with large $\tilde{m}$)
have the usual sign of the coupling (a decrease of optical frequency
on expansion), while those with small $\tilde{m}$ have the opposite
sign. In a ray-optical picture, they travel along great circles passing
near the pole, and feel an overall reduction of path length when the
drop's equator expands.

The preceding discussion applies strictly to a perfectly spherical
drop. In practice, the magnetic fields used to counteract the pull
of gravity tend to distort the drop's shape \cite{hill_vibrations_2010}.
A rotating drop will also experience distortion due to centrifugal
forces. Such distortions break the degeneracy of the optical WGMs.
Eq.~(\ref{eq:GeneralCouplingConstant}), with $\delta R_{{\rm ZPF}}$
replaced by the change of radius $\delta R$, can also be used to
estimate the impact of this distortion on the optical WGMs. A family
of modes with any given $\tilde{l}$ splits into $\tilde{l}+1$ distinct
frequencies (as modes with given $|\tilde{m|}$ remain degenerate),
with the frequency shift $\propto\tilde{m}^{2}$. In the case of modes
with $\lambda=1\ \mu{\rm m}$, $R=1{\rm \ mm}$ scenario and a distortion
$\delta R/R\sim1\%$, the originally degenerate multiplet would split
into a band with $\sim$THz bandwidth, far larger than the vibrational
frequencies we consider. Indeed, the bandwidth of frequencies produced
from each $\tilde{l}$ manifold would exceed the free spectral range
of the WGMs by more than an order of magnitude, meaning that optical
modes with differing $\tilde{l}$ could undergo avoided crossings
for certain values of the distortion.

\section{Mechanical and Optical Quality Factors}

\subsection{Damping of mechanical modes}

As described in the introduction, the combination of superfluidity
and magnetic levitation should strongly suppress some sources of mechanical
damping. Here we consider the two mechanisms which are expected to
dominate the energy loss from the mechanical modes of a $^{4}\mathrm{He}$
drop. The first is due to damping by the He gas surrounding the drop,
and the second is the exchange of mechanical energy between the drop's
mechanical modes (i.e., mediated by its mechanical nonlinearity).
Both of these processes are strongly temperature-dependent.

\begin{figure}
\begin{centering}
\includegraphics[width=0.8\columnwidth]{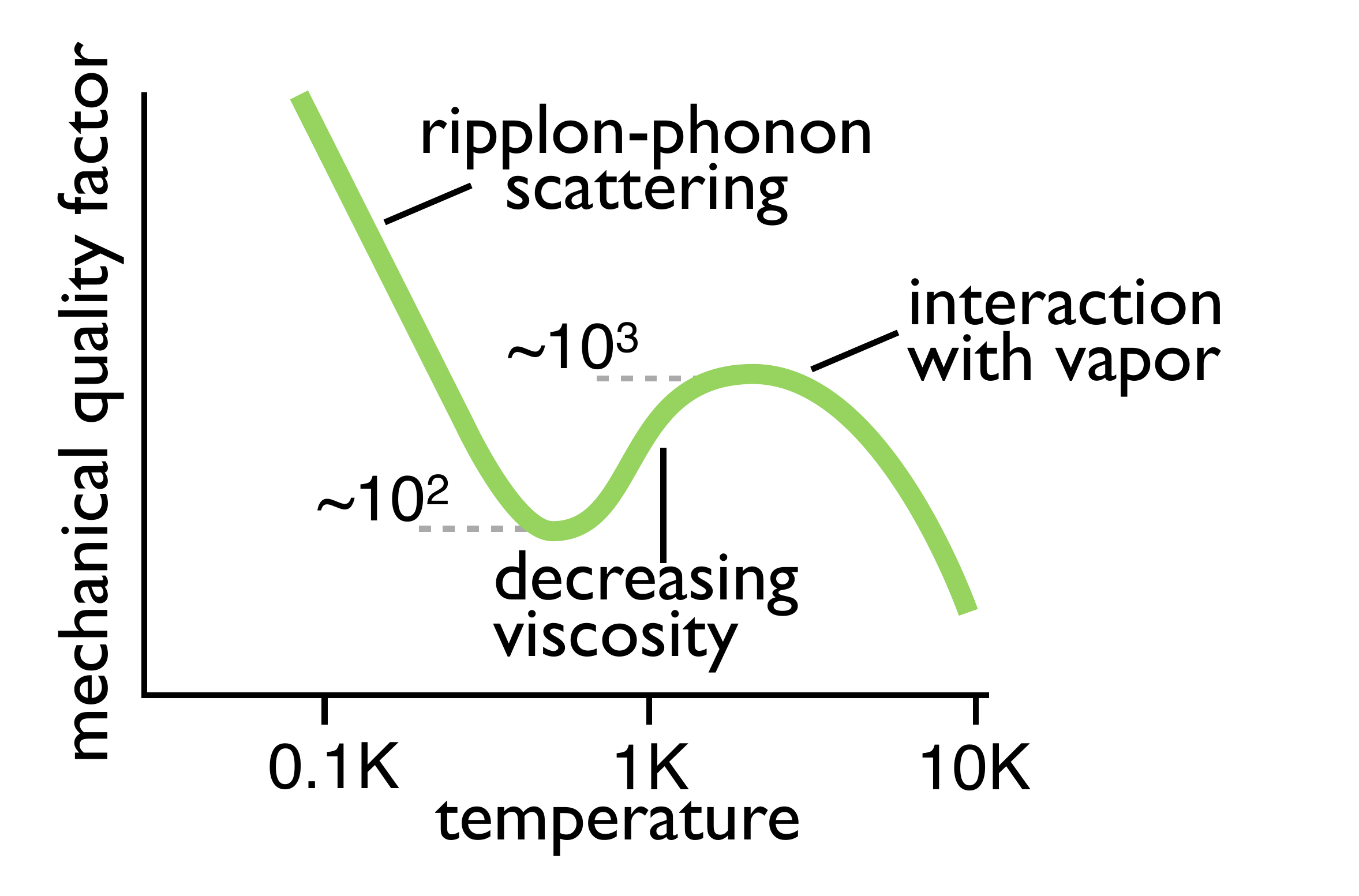}
\par\end{centering}
\caption{\label{Figure_MechanicalDamping}Qualitative sketch of the expected
temperature dependence for the mechanical quality factor $Q_{\mathrm{{mech}}}$
of $l=2$ surface modes in a $R=1\ {\rm mm}$ $^{4}{\rm He}$ drop,
with indication of different regimes. In the regime of viscous damping,
the viscosity of the normal component \emph{drops} with increasing
temperature, which leads to nonmonotonic behavior of the quality factor.}
\end{figure}

At sufficiently high temperatures, the vapor surrounding the drop
and the thermal excitations within the drop are dense enough to be
described as hydrodynamic fluids. Experiments in this regime measured
the quality factor $Q_{\mathrm{{mech}}}$ of the $l=2$ surface modes
for a $^{4}{\rm He}$ drop of radius $R=2$ mm for $0.65$ K$\leq T\leq1.55$
K \cite{whitaker_shape_1998}. The measured $Q_{\mathrm{{mech}}}(T)$
was in good agreement with calculations based on a hydrodynamic treatment
of the three fluids (i.e., the superfluid, normal fluid, and vapor)
\cite{whitaker_theory_1999}. Within this temperature range, $Q_{\mathrm{{mech}}}$
reached a maximum value ($\sim1200$) for $T\sim1.2$ K (see Fig.
\ref{Figure_MechanicalDamping}). At higher $T$, the decrease in
$Q_{\mathrm{{mech}}}$ is due to the higher vapor density. At lower
$T$ the decrease in $Q_{\mathrm{{mech}}}$ is due to the increasing
dynamic viscosity of $^{4}{\rm He}$. 

The counterintuitive increase in viscosity with decreasing $T$ reflects
the increasing mean free path $\Lambda$ of the thermal phonons within
the drop. Since $\Lambda$ is proportional to $T^{-4}$ \cite{whitworth_experiments_1958,maris_hydrodynamics_1973},
at still lower temperatures the drop will enter a new regime in which
$\Lambda>R$. In this regime the hydrodynamic description fails and
$Q_{\mathrm{{mech}}}$ is expected to increase again. Some support
for this picture can be found in the measurements of Ref. \cite{webeler_viscosity_1965,webeler_nasa_1968}.
For a $R=1$ mm drop, this regime should occur for $T<0.4\:K$.

At these low temperatures, the dominant loss mechanism for the surface
waves (ripplons) is ripplon-phonon-phonon scattering, in which a thermally
excited bulk phonon scatters off the ripplon and is Doppler-shifted,
carrying away energy. This effect has been studied experimentally
and theoretically in \cite{roche_interpretation_1996}, with a resulting
estimate for the $Q_{\mathrm{{mech}}}$ of a surface wave traveling
on a plane surface:

\begin{equation}
\frac{1}{Q_{\mathrm{{mech}}}}=\frac{\pi^{2}}{90}\frac{\hbar k}{\rho\omega}\left(\frac{k_{B}T}{\hbar v_{s}}\right)^{4}\,.
\end{equation}
Here $\omega$ is the surface mode frequency, $k$ is its wavenumber,
$\rho$is the density, and $v_{s}$ is the sound velocity.

While our proposal focuses on mechanical modes of $^{4}\mathrm{He}$
drops, for completeness we also note the mechanical losses of $^{3}\mathrm{He}$
drop surface modes. For a normal-fluid $^{3}{\rm He}$ drop, one can
apply Chandrasekhar's result for the viscous damping \cite{chandrasekhar_oscillations_1959},
according to which $1/Q_{\mathrm{{mech}}}=\mu(l-1)(2l+1)/(\omega R^{2}\rho)$.
Here $\mu$ is the dynamic viscosity and $\rho$ is the density. For
$T=$ 1 K, where $\mu=30\ \mu$P, a 1 mm drop would have $l=2$ surface
modes with $Q_{\mathrm{{mech}}}\approx70$, and the quality factor
would decrease approximately as $Q_{\mathrm{{mech}}}\propto T^{2}$
at lower temperatures \cite{betts_viscosity_1963,betts_viscosity_1965}.
For $T\lesssim1$ mK, a $^{3}\mathrm{He}$ drop would become superfluid;
however this temperature range is not likely to be accessed via the
cooling methods considered here.

\subsection{Damping of optical whispering gallery modes}

Light confined within a WGM may experience loss due to radiation from
the evanescent portion of the mode, scattering from surface roughness
or bulk defects, or absorption by the host material or its impurities
\cite{oraevsky_whispering-gallery_2002}. Here we consider the contributions
of each of these mechanisms to the quality factor of the optical WGMs
in a levitated drop of liquid helium.

Optical WGMs have been studied in drops of several different types
of liquid. Pioneering experiments by the Chang group \cite{tzeng_laser_1984,qian_lasing_1986,popp_q_1997}
focused on WGMs in freely-falling drops of ethanol and water and found
optical $Q_{\mathrm{{opt}}}$ as high as $10^{8}$. Measurements of
WGMs in suspended drops of oil show $Q_{\mathrm{{opt}}}=1.7\times10^{8}$
\cite{dahan_droplet_2016}. Pendant drops of cryogenic liquid $\mathrm{H}_{2}$
\cite{uetake_stimulated_1999,uetake_stimulated_2002} demonstrated
$Q_{\mathrm{{opt}}}=4.2\times10^{9}$.

In comparison with these materials, liquid ${\rm He}$ should offer
reduced absorption. This is because ${\rm He}$ is monoatomic (removing
the possibility of inelastic light scattering from bond stretching
or other molecular degrees of freedom), has a large gap for electronic
excitations ($\sim$19 eV), and is free of chemical impurities and
surface adsorbates.

Liquid He posesses an unusally low index of refraction ($n\sim1.028$),
which would lead to increased radiative loss at fixed $R$ and $\lambda$.
However radiative loss from a spherical resonator decreases exponentially
\cite{oraevsky_whispering-gallery_2002} with $R/\lambda$. As a result,
even with the small refractive index of He, radiative loss becomes
negligible in mm-scale drops (see Fig. \ref{Fig3}). 

Surfaces defined by surface tension are typically very smooth. Nevertheless,
thermally excited ripplons will result in an effective surface roughness.
As described below, we expect this will be the dominant loss mechanism.
To analyze this mechanism we assume that the random thermal surface
deformation is essentially frozen during the lifetime of the optical
WGM. Furthermore, we only consider ripplon modes with wavelengths
small compared to $R$. In this case the Fourier transform $\tilde{G}(k)$
of the spatial correlation function of surface deflections can be
approximated by the known result for a planar surface, $\tilde{G}(k)=2\pi k_{B}T/\sigma\left|k\right|$,
where $\sigma$ is the surface tension. Adapting an analysis for planar
waveguides with a disordered surface \cite{lacey_radiation_1990},
the WGM loss rate (via outscattering) is

\begin{equation}
\frac{1}{Q_{{\rm opt}}}\approx\Phi(0)^{2}(\epsilon-1)^{2}\frac{k_{0}^{2}}{8\pi}\int_{0}^{\pi}\tilde{G}(k-k_{0}\cos\theta)d\theta\,.
\end{equation}
Here $k_{0}$ is the optical WGM's vacuum wavenumber, and $\epsilon=1.057$
is the dielectric constant of helium. $\Phi(y)$ is the normalized
transverse mode shape ($\int\Phi(y)^{2}dy=1$), such that $\Phi(0)^{2}$,
evaluated at the surface, is roughly the inverse extent of the mode.
Following Ref. \cite{gorodetsky_rayleigh_2000} and considering TE
modes only, we take $\Phi(0)^{2}\approx2\epsilon/(R(\epsilon-1))$
as an upper estimate, eventually obtaining $Q_{{\rm opt}}\approx2R/(\pi k_{0}\sqrt{\epsilon-1})(\sigma/k_{B}T)$
as a lower bound for $Q_{{\rm opt}}$. Applying this approach to liquid
$^{4}{\rm He}$ at $T=300\ {\rm mK}$, with $\sigma=3.75\cdot10^{-4}$
N/m, and $\lambda=1\ \mu{\rm m}$ gives $Q_{\mathrm{{opt}}}\sim4\cdot10^{10}$
for a drop with $R=1\ \mathrm{mm}$. For $^{3}\mathrm{{He}}$, the
surface tension and the resulting $Q$ are both about 2.5 times lower.

At present there are no experiments on He drops with which to compare
this estimate. However applying this analysis to the liquid $\mathrm{H}_{2}$
drops of Refs. \cite{uetake_stimulated_1999,uetake_stimulated_2002},
gives $Q_{\mathrm{{opt}}}\sim2\cdot10^{8}$, i.e. it underestimates
$Q_{\mathrm{{opt}}}$ by roughly an order of magnitude. This may reflect
the fact that the ripplon modes evolve during the WGM lifetime, averaging
out some of the effective roughness. 

\textcolor{red}{}

We estimate other scattering mechanisms to be significantly less important:
Brillouin scattering from thermal density fluctuations inside the
drop \cite{seidel_rayleigh_2002,weilert_laser_1995} should give $Q_{\mathrm{{opt}}}$
> $10^{13}$, and Raman scattering from rotons should be even weaker
(following Ref. \cite{greytak_light_1969}). 

\subsection{Summary of parameters}

Based on the estimates above, the most important optomechanical parameters
for a drop of $^{4}{\rm He}$ with $R=1\:{\rm mm}$ are summarized
in the following table (assuming $T=300\,$ mK):

\begin{center}
\begin{tabular}{cccc}
\toprule 
$\omega_{{\rm vib}}/2\pi$ & $Q_{{\rm mech}}$ & $Q_{{\rm opt}}$ & $g_{0}/2\pi$\tabularnewline
\midrule
\midrule 
$23{\rm \,Hz}$ & $>10^{3}$ & $>10^{10}$ & $213{\rm \,Hz}$\tabularnewline
\bottomrule
\end{tabular}
\par\end{center}

Notably, this system enters the previously-unexplored regime where
$g_{0}>\omega_{\mathrm{{vib}}}$. While our estimate for $Q_{\mathrm{{opt}}}$
gives an optical linewidth that is only $\sim$40 times larger than
the optomechanical coupling rate, the same ``frozen-deformation''
approximation underestimates the quality factor of hydrogen drops
by a factor of 20. Moreover, at lower temperatures, $Q_{\mathrm{{opt}}}\propto1/T$
increases yet further. The levitated helium drop is thus likely to
approach the single-photon strong coupling regime. 

\subsection{Evaporative Cooling}

The temperature of an optomechanical device is typically set by the
competition between optical absorption (which leads to heating) and
the device's coupling to a thermal bath (which allows this heat to
be removed). For levitated solids, the heat removal process is inefficient,
as it occurs primarily via blackbody radiation, resulting in elevated
temperatures for even moderate optical power. In contrast, a levitated
liquid may also cool itself via evaporation. As described below, evaporation
provides an effective means for maintaining the drop temperature well
below 1 K. However evaporation also couples the drop's radius $R$
to its temperature $T$. Since many of the device's relevant parameters
(such as the resonance frequencies and quality factors of the optical
and mechanical modes) depend on both $R$ and $T$ it is important
to have a quantitative model of the evaporation process.

Evaporative cooling of helium droplets has been studied both experimentally
and theoretically. Experiments to date have used $\mu$m- and nm-scale
droplets that are injected into a vacuum chamber. In the $\sim$ms
time before the droplets collide with the end of the vacuum chamber
they are found \cite{toennies_superfluid_2004} to reach $T\sim370$
mK (150 mK) for $^{4}$He ($^{3}$He). This cooling process can be
understood by considering how energy loss \textendash{} given by the
latent heat per atom ($\Delta E(T)$) times the evaporation rate $\Gamma(N,T)$
(atoms/sec) \textendash{} leads to cooling according to the heat capacity
$C(N,T)$ of the droplet: $\frac{dT}{dt}=-\Gamma(N,T)\Delta E(T)\frac{1}{C(N,T)}$,
where the total number $N$ of atoms in the drop decreases as $\frac{dN}{dt}=-\Gamma(N,T).$
Simultaneous solution of the differential equations yields the cooling
dynamics. Theoretical models valid in the low-$T$, low-$N$ limit
have successfully explained the experiments \cite{brink_density_1990}.
They used an Arrhenius law for the evaporation rate $\Gamma\propto NT^{2}e^{-E_{0}/k_{B}T}$with
$E_{0}=\Delta E(0)=k_{B}*7.14\:{\rm K}\:(2.5\:\mathrm{K})$ for $^{4}{\rm He}\:(^{3}{\rm He})$,
and considered only ripplon (for $^{4}{\rm He}$) or free Fermi gas
(for $^{3}{\rm He}$) contributions to the heat capacity of the drop. 

\begin{figure}
\includegraphics[width=0.8\columnwidth]{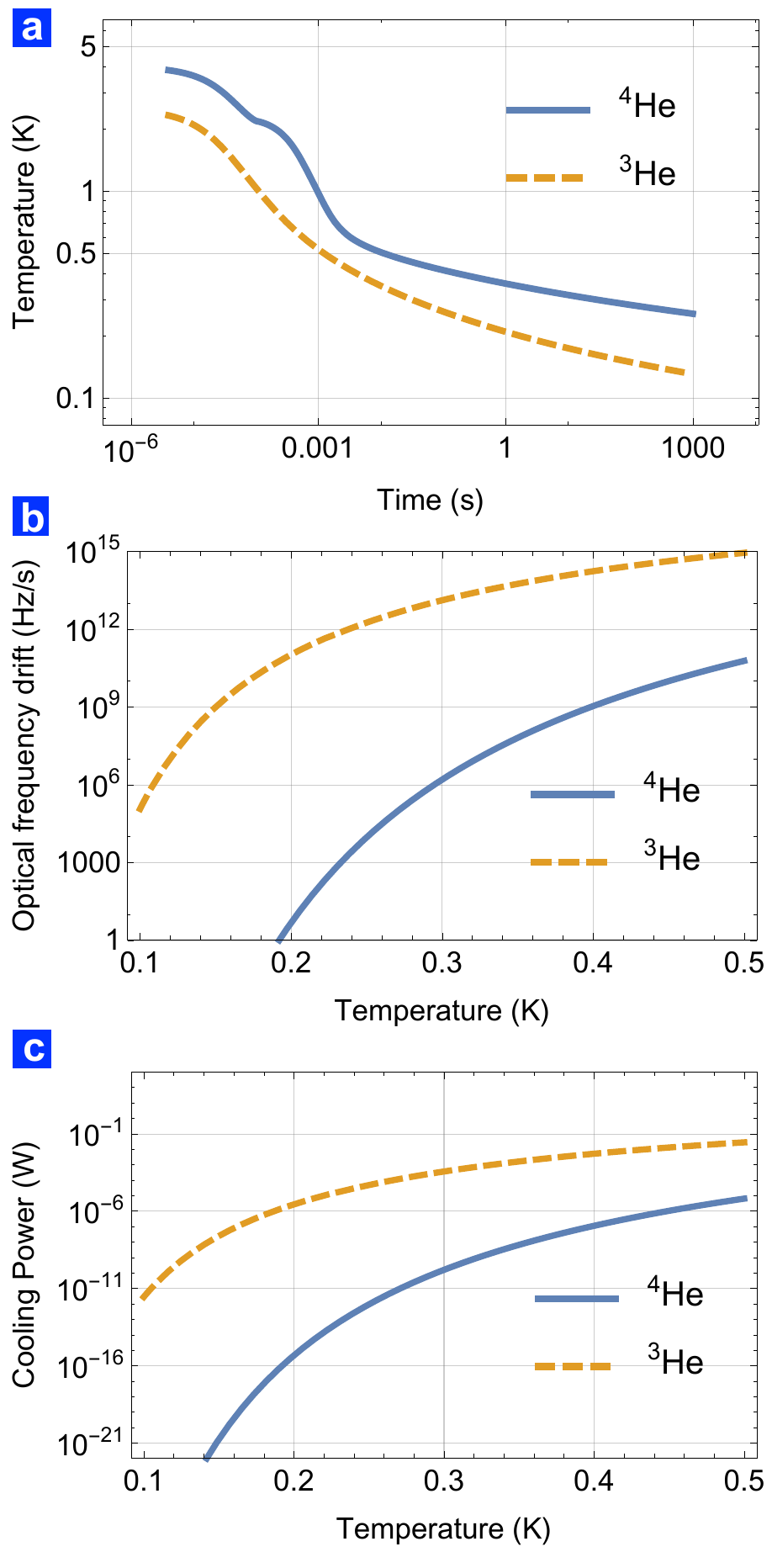}

\caption{\label{Fig_EvaporativeCooling}Evaporative cooling of the helium drop.
(a) Evolution of temperature as a function of time for drops with
an initial radius $R=1$ mm (which decreases by about 10\% during
cooling). Note the logarithmic time scale; the physically relevant
times are those above $1\,{\rm sec}$, lower times depend on the detailed
experimental protocol. (b) Rate of change of the whispering gallery
mode resonance frequency, due to the decrease of radius by continuous
evaporation, for a $\lambda=1\:{\rm \mu m}$ mode of a $R=1{\rm \:mm}$
drop. (c) Cooling power $\Delta E\,\Gamma$, displayed as a function
of temperature, for $R=1\:{\rm mm}$. Blue: $^{4}{\rm He}$; Dashed
orange: $^{3}{\rm He}$.}
\end{figure}

To model the full range of temperatures attained during cooling, and
to account for phonon contributions to the heat capacity (needed for
large-$N$ drops of $^{4}{\rm He}$), we use primarily measurement-based
values \footnote{For $^{4}{\rm He}$ vapor pressure at temperatures below 0.65K, we
use the theoretical Arrhenius law for evaporation rate \cite{brink_density_1990},
which fits experimental data well even at more elevated temperatures.
All other values are found by interpolating published experimental
data or using published empirical formulas within their region of
validity.} of latent heat $\Delta E$, vapor pressure $P$ (which determines
the evaporation rate via $\Gamma\approx4\pi R^{2}P/\sqrt{2\pi mk_{B}T}$
assuming unit accommodation coefficient), and specific heat \cite{donnelly_observed_1998,greywall_specific_1983,kerr_orthobaric_1954,huang_practical_2006}.
Figure \ref{Fig_EvaporativeCooling}a shows the expected temperature
$T(t)$ for $^{4}{\rm He}$ and $^{3}{\rm He}$ drops with an initial
radius of 1 mm, cooled from 4.0 K and 2.5 K respectively. Because
$^{3}{\rm He}$ has a higher vapor pressure, it cools more effectively:
For $^{4}{\rm He}$ ($^{3}{\rm He}$), the drop temperature reaches
$\sim$350 mK ($\sim$200 mK) after $\sim$ 1 s evaporation time and
slowly cools to $\sim$290 mK ($\sim$150 mK) after $\sim$ 1 minute.
The complete cooling process shrinks the radius of both types of drops
by about 10\%.

In the absence of any heat load (as assumed for the simulation shown
in Fig. \ref{Fig_EvaporativeCooling}a), $T$ will continue to decrease,
although over impractically long time scales. In an actual experiment
we expect a finite heat load on the drop, which will result in $T$
asymptoting to a finite value. The asymptotic value of $T$ will determine
the quality factor of the optical and mechanical modes (as described
above). It will also set the (constant) rate at which $R$ will drift
during any experiment. This drift in $R$ will not result in any appreciable
change in the mechanical mode frequencies; however the drift in the
optical mode frequency will need to be tracked, e.g., by standard
laser-locking techniques (see Fig. \ref{Fig_EvaporativeCooling}b,c).

For a $^{4}{\rm He}$ drop with $R=$ 1 mm, the optical drift rate
is $\sim10^{16}\ \mathrm{Hz/s}$ per Watt of dissipated power (and
is $\sim4\times$ larger for $^{3}{\rm He}$ because of the lower
binding energy and density of $^{3}{\rm He}$). To estimate the likely
heatload on the drop, we note that Brillouin scattering in the optical
WGM \cite{weilert_magnetic_1996} should result in absorption of $<10^{-10}$
of the incident laser power (for $\lambda=1\ \mu$m). Assuming an
input power $\sim\mu$W, this would result in an optical drift rate
of only $\sim$Hz/s.

\section{Rotations}

\subsection{Towards Quantum Non-Demolition Measurements of Rotation}

One of the unique characteristics of fluid drops, as opposed to solid
dielectric spheres, is the possibility to optically measure and possibly
even control rotations, via the deformation of the rotating drop.
Rotational motion represents a low-energy excitation that is not equivalent
to a harmonic oscillator, and so offers access to quantum phenomena
that are qualitatively distinct from those typically studied in cavity
optomechanics.

The rotational motion of $^{4}{\rm He}$ is qualitatively different
from that of $^{3}{\rm He}$. For the temperatures relevant here ($\sim300{\rm \,mK}$),
$^{4}{\rm He}$ is a pure superfluid and so its rotation is determined
by the presence of vortices, each with quantized circulation. The
angular momentum associated with each vortex is $N\hbar$ (where $N$
is the number of atoms in the drop); thus the drop's angular momentum
can only change in relatively large discrete steps. In practice, this
will ensure that the number of vortices is constant at low temperatures.
Nevertheless, a drop a with a fixed number of vortices will still
possess nontrivial dynamics owing to the vortex lines' motion.

In contrast, $^{3}{\rm He}$ is a normal fluid at these temperatures
and so may undergo rigid-body rotation. Its angular momentum can change
in very small steps of $\hbar$, allowing the drop's total angular
momentum to be a dynamical variable. Although $^{3}{\rm He}$ is highly
viscous at these temperatures, viscosity does not directly damp rigid
body rotation.

For both $^{4}{\rm He}$ and $^{3}{\rm He}$, the drop's rotational
motion is expected to interact with the optical WGMs primarily because
the flow field associated with the rotation will deform the drop shape,
and thereby detune the WGMs. This coupling would allow optical measurements
(i.e., of the WGM) to provide information about the drop's rotational
motion. In order to consider the quantum limits of such a measurement,
we note that the angular momentum $L_{z}=I\Omega_{z}$ is connected
to the angular frequency $\Omega_{z}$ via the drop's moment of inertia
$I=(8\pi/15)\rho R^{5}$ (here we assume that the drop is nearly spherical).
In principle, $\Omega_{z}$ can be inferred from the WGM detuning
caused by the equatorial bulge (which is produced by the centrifugal
acceleration $\Omega_{z}^{2}R$). The radius at the equator increases
by an amount $\delta R$ $\propto\Omega_{z}^{2}$. As described above,
the resulting shift of an optical WGM at the equator is $\delta\omega_{\mathrm{opt}}=\omega_{{\rm opt}}\delta R/R$.
We thus obtain an ``opto-rotational'' coupling Hamiltonian of the
form 

\begin{equation}
\hat{H}_{{\rm QND}}=\hbar g_{L}\left(\frac{\hat{L}_{z}}{\hbar}\right)^{2}\hat{a}^{\dagger}\hat{a}\,.\label{eq:QNDHamiltonian}
\end{equation}
The form of this Hamiltonian allows for a QND measurement of $\hat{L}_{z}^{2}$. 

The Hamiltonian of Eq.~(\ref{eq:QNDHamiltonian}) is a simplified
version of the real coupling, as will be explained in the next section.
However, it is sufficient for understanding the basic physics of the
opto-rotational coupling, and to estimate the feasibility of angular
momentum QND measurements.

The frequency shift $g_{L}$ in Eq.~(\ref{eq:QNDHamiltonian}) is
given by $g_{L}=\omega_{{\rm opt}}(\delta R/R)\left(\hbar/L_{z}\right)^{2}$,
where $L_{z}$ is the (classical) mean value of the drop's angular
momentum and $\delta R$ is the bulge produced by $L_{z}$. By balancing
pressure, centrifugal force, and surface tension we find: 

\begin{equation}
\delta R=(\rho/\sigma)R^{4}\Omega_{z}^{2}/24\,.\label{BulgeFromRotation}
\end{equation}
Thus, smaller drops deform less for a given angular frequency, due
to the smaller centrifugal force. However, in terms of $g_{L}$ this
is overcompensated by the rapidly increasing ratio $\Omega_{z}/L_{z}=1/I$.
Altogether, the WGM detuning has a strong dependence on the drop radius: 

\begin{equation}
g_{L}=\omega_{{\rm opt}}\frac{\hbar^{2}}{\rho\sigma R^{7}}\frac{1}{24}\left(\frac{15}{8\pi}\right)^{2}
\end{equation}
Nevertheless, it should be stressed that for typical parameters this
constant is exceedingly small. For a $^{3}{\rm He}$ drop with $R=1\ {\rm mm}$,
$\rho\sim81\ \mathrm{\mathrm{kg/}m^{3}}$, and $\sigma=1.52\cdot10^{-4}\mathrm{\ N/m}$,
we have $g_{L}=\omega_{{\rm opt}}\cdot1.3\cdot10^{-47}$. Fortunately,
in most situations the detuning can be much larger than that. This
is because the WGM detuning scales with $\hat{L}_{z}^{2}$, meaning
that changing $L_{z}$ by $\hbar$ results in a detuning $2g_{L}(L_{z}/\hbar)$
and so can be substantially enhanced for large values of $L_{z}/\hbar$. 

In order to detect a given deviation in angular momentum $\delta L_{z}$,
a phase shift $\sim\delta\omega_{{\rm opt}}/\kappa$\textbf{ }has
to be resolved by the number of photons $N_{{\rm phot}}$ sent through
the drop's WGM during the time of the measurement. This implies that
the minimum detectable phase must be sufficiently small, $\delta\theta=1/(2\sqrt{N_{{\rm phot}}})<\delta\omega_{{\rm opt}}/\kappa=Q_{{\rm opt}}\delta\omega_{{\rm opt}}/\omega_{{\rm opt}}$.
More formally, the resolution is set by $\delta L_{z}^{2}=t_{{\rm meas}}^{-1}\cdot S_{L}$,
where we have introduced the spectral density $S_{L}$ for the angular
momentum imprecision noise. The spectral density is defined in the
usual way \cite{clerk_introduction_2010}, with $S_{L}=\int\left\langle \delta L_{z}(t)\delta L_{z}(0)\right\rangle dt$,
where $\delta L_{z}(t)$ represents the instantaneous fluctuations
of the angular momentum deduced from the observed phase shift. Taking
into account the phase-shift fluctuations produced by the shot-noise
of the laser beam, as estimated above, we find:

\begin{equation}
S_{L}\equiv\hbar^{2}\left(\frac{\omega_{{\rm opt}}}{2g_{L}}\right)^{2}\left(\frac{\hbar}{L_{z}}\right)^{2}(4Q_{{\rm opt}}^{2}\dot{N}_{{\rm phot}})^{-1}\,.
\end{equation}

We briefly discuss a numerical example to illustrate the possible
experimental measurement precision. A normal $^{3}\mathrm{{He}}$
drop spinning at $\Omega_{z}/2\pi=1$ Hz (well below the hydrodynamic
instability) will have $L_{z}/\hbar=I\Omega_{z}/\hbar=8\cdot10^{21}$.
For $\omega_{{\rm opt}}/2\pi=300\,{\rm THz}$ $(\lambda=1\ \mu\mathrm{{m}})$,
this yields an optical frequency shift of $2g_{L}(L_{z}/\hbar)\approx2\pi\cdot6\cdot10^{-11}{\rm \ Hz}$
per $\hbar$ of additional angular momentum. Therefore, we find $\sqrt{S_{L}}\approx2\cdot10^{24}\hbar/(Q_{{\rm opt}}\sqrt{\dot{N}_{{\rm phot}}})$.
For $10\ {\rm \mu W}$ of input power and for $Q_{{\rm opt}}=10^{10}$,
one would thus have an angular momentum resolution of $\sqrt{S_{L}}\approx3\cdot10^{7}\hbar/\sqrt{{\rm Hz}}$.

These numbers indicate that it will be impossible to resolve a change
of angular momentum by a single quantum $\hbar$. However, one should
be able to measure $L_{z}$ (or $L_{x}$ or $L_{y}$) with a precision
better than $\sqrt{\hbar L}$. This is the spread of $L_{x}$ and
$L_{y}$ in a situation with maximum $L_{z}=L$, according to Heisenberg's
uncertainty relation. Indeed, for the example given above, $\sqrt{\hbar L}\sim10^{11}\hbar$,
which, according to the estimated noise power $S_{L}$, can be resolved
in $t_{{\rm meas}}\sim0.1\:\mu s$. Moreover, in the case of a superfluid
$^{4}\mathrm{{He}}$ drop under identical conditions, the sensitivity
$\sqrt{S_{L}}\approx1.4\cdot10^{8}\hbar/\sqrt{{\rm Hz}}$ is easily
sufficient to carefully monitor a single vortex line, which would
carry an angular momentum of $\sim10^{20}\hbar$

There are three potential noise sources that may interfere with the
QND measurement of angular momentum: fluctuations in the number of
evaporating atoms leading to stochastic changes of the drop radius,
random angular momentum kicks due to evaporating atoms, and angular
momentum transfer by randomly out-scattered photons. We have estimated
all these effects (see Appendix A), and found them to be smaller than
the measurement uncertainty attained in the example given above.

Lastly, we note that in addition to the centrifugal coupling considered
above there is also the Fizeau effect, which produces a WGM detuning
$\propto\hat{L}_{z}$ (with a different sign for clockwise and counter-clockwise
WGM modes). We estimate the single quantum coupling rate for the effect
to be $g_{F}\approx2\pi\times10^{-20}$ Hz for $R=1\:{\rm mm}$. Since
the Fizeau effect does not increase with $|L_{z}|$, we expect the
centrifugal coupling to dominate.

\subsection{Coupling between Vibrations and Rotations}

The coupling in Eq.~(\ref{eq:QNDHamiltonian}) is idealized in two
ways. First, it assumes that the drop strictly rotates only around
the $z$-axis and that $\hat{L}_{x},\,\hat{L}_{y}$ are not involved
in the dynamics. Second, we have written down a direct coupling between
rotation and optical frequency. In reality, the rotation will first
lead to a deformation, i.e. a displacement of one of the surface modes,
and this deformation will then couple to the optical WGM. Conversely,
the laser's shot noise will lead to a fluctuating force acting on
the surface modes, which then couple back to the rotation. This represents
the back-action associated with the optical readout. 

In the present QND case, the back-action leads to dephasing between
different eigenstates of the angular momentum projection $\hat{L}_{z}$.
Physically, fluctuations in the circulating photon number couple to
$\hat{L}_{z}^{2}$ (via the deformation) which then scramble $\hat{L}_{x}$
and $\hat{L}_{y}$. 

\begin{figure}
\includegraphics[width=1\columnwidth]{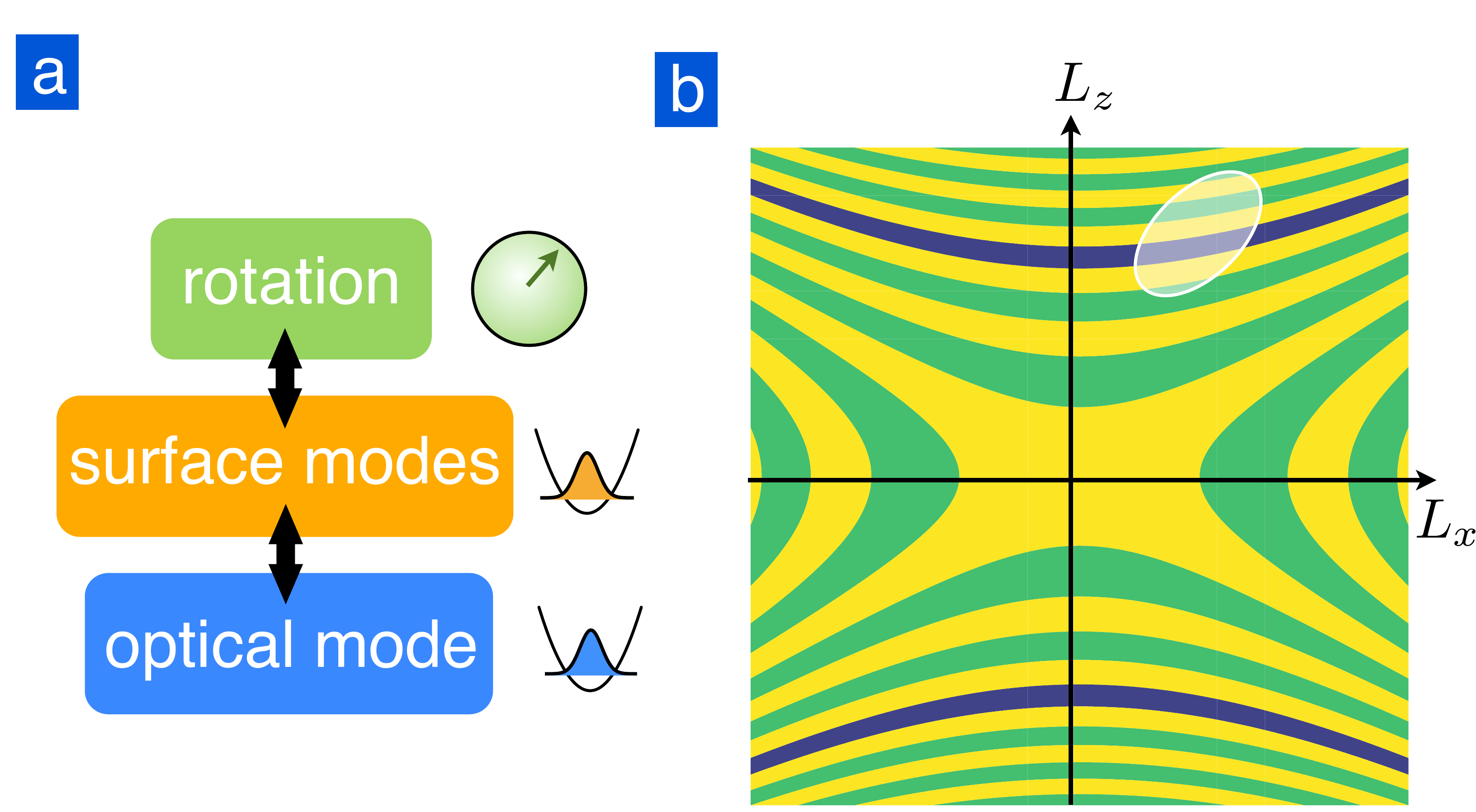}

\caption{\label{FigMeasurement}(a) The helium drop represents a novel coupled
system, with nonlinear interactions leading from rotational motion
via the surface modes to optical frequency shifts. (b) A measurement
of the WGM detuning reveals information about $L_{z}^{2}-(L_{x}^{2}+L_{y}^{2})/3$.
Contours of this function are shown in the $L_{x,}L_{z}$ plane. A
given measurement result (within some uncertainty interval) maps to
a narrow region (dark blue areas), whose intersection with the state's
initial uncertainty (white ellipse) determines the state after the
measurement.}

\end{figure}

In summary, a more complete understanding of the optical measurement
of angular momentum will require a description of the coupling between
mechanical vibrations and the drop's rotations. This is also an interesting
dynamical problem in its own right, and it turns the liquid drop into
a novel coupled opto-mechanical-rotational system (Fig.~\ref{FigMeasurement}a). 

The interplay between rotations, deformations and vibrations in fluid
spheres has been studied in nuclear physics (for the liquid drop model
of the nucleus \cite{bohr_nuclear_1998}), geophysics (for rotating
planets), and hydrodynamics (for rotating drops \cite{busse_oscillations_1984}).
For small angular frequencies, the two most important effects are
(i) the slight deformation of the drop due to the centrifugal force
and (ii) a shift in the frequencies of the surface modes. This frequency
shift (sometimes known as Bryan's effect \cite{bryan_onthe_1890,joubert_rotating_2009})
is due to the Coriolis force. It leads to a rotation of the surface
vibrations that is neither a simple co-rotation with the rotating
drop nor static in the lab frame. For the $l=2$ modes of interest
here, the frequencies in the rotating frame are shifted by $-\omega_{{\rm rot}}m/2$,
where $m$ is the mode index ($\left|m\right|\leq2$).

Previous studies of the interplay of rotations and vibrations have
typically been limited to a fixed rotation axis or other special cases
\cite{bohr_nuclear_1998,busse_oscillations_1984}. To move beyond
these assumptions, we have derived the full Lagrangian of the system
without any such assumptions of symmetry, for the case where only
$l=2$ surface modes are excited (extensions to larger $l$ are straightforward).
To accomplish this, we note that the surface deformation pattern $\delta R(\theta,\varphi,t)$
in the laboratory frame can be decomposed into spherical harmonics.
The five deflection amplitudes $X_{m}$ of the $l=2$ surface modes,
together with the three Euler rotation angles, form the set of variables
in the Lagrangian (Appendix B). 

The Lagrangian can be derived by (i) calculating the flow field inside
the drop enforced by the time-varying deformation pattern of its surface,
(ii) integrating the resulting kinetic energy density over the volume
of the drop, and (iii) adding the potential energy from the surface
tension. This assumes an incompressible fluid whose flow field can
be understood as an irrotational flow pattern in the co-rotating frame,
produced by the surface deformation. The final result involves the
deformation variables $X_{m}$, the angular velocity vector $\mathbf{\Omega}$,
and the Euler angles that transform between the co-rotating frame
and the lab frame. We display the slightly involved Lagrangian in
the appendix and we will publish its full derivation elsewhere.

The basic physics can be understood qualitatively by considering the
special case of a rotation around the $z$-axis. In particular, the
kinetic energy in the Lagrangian contains the following term, beyond
the standard terms for the rigid-body rotation of a sphere and the
kinetic energies of the surface modes:

\begin{equation}
\frac{I}{4}\sqrt{\frac{5}{\pi}}\frac{X_{0}}{R}\Omega_{z}^{2}.
\end{equation}
This is the term that couples the bulge mode deflection $X_{0}$ to
the rotation around the $z$-axis (with the moment of inertia $I=(8\pi/15)\rho R^{5}$).
Physically, it can be read in two ways. First, spinning up the drop
creates a finite deflection proportional to $\Omega_{z}^{2}$, which
then leads to an optical shift, as discussed previously. Conversely,
a deflection increases the moment of inertia and thereby the rotational
energy for a given angular frequency. 

We note that for a rotating drop there also appears a set of low-frequency
modes, the so-called ``inertial modes'' \cite{greenspan_theory_1968,landau_fluid_2013}.
Their frequencies scale with the rotation frequency, and they are
thus well separated from the vibrational modes we have been discussing,
as long as the rotation speed is sufficiently far below the instability
threshold for nonlinear drop deformation and fission. As a result,
we neglect them.

As for the effective coupling between the angular momentum and the
optical frequency, we have to point out another interesting aspect
that has been omitted in the simplified model of Eq.~(\ref{eq:QNDHamiltonian}).
An optical whispering gallery mode traveling around the equator in
the $xy$-plane will be sensitive not only to the bulge equatorial
deformation that is generated by $L_{z}$; its frequency will also
be shifted by a rotation around the x-axis (or y-axis), since this
leads to an expansion of the equator in the $yz$- (or $xz$-) plane.
According to Eq.~(\ref{eq:GeneralCouplingConstant}), this frequency
shift is $1/3$ of that obtained for $z$-rotations, and has the opposite
sign. As a consequence, the operator that is really measured is expected
to be the combination $\hat{L}_{z}^{2}-\frac{1}{3}(\hat{L}_{x}^{2}+\hat{L}_{y}^{2})$.
The situation is displayed in Fig.~\ref{FigMeasurement}b.

In an experiment, angular momentum will be generated by spinning up
the drop (e.g. via the application of a rotating electric field).
Such an approach will not select a single energy eigenstate with a
definite $L$, but rather a coherent superposition of various $L$
(as well as of various $L_{z}$). The details will depend on the exact
procedure used for spinning up the drop, and in practice there will
be a thermal incoherent mixture because the experiment is conducted
at finite $T$ with a large thermal population of vibrational and
rotational levels. The QND measurements described above would then
be able to resolve the angular momentum to some extent, thereby narrowing
its distribution via the measurement backaction.

In summary, the Lagrangian that we briefly discussed here will form
the general basis for discussions of the intricate coupled nonlinear
dynamics of vibrations and rotations in the fluid drop. Among other
things, this will enable a detailed analysis of the measurement backaction
in optical dispersive measurements of the angular momentum components.
However, exploring the rich nonlinear dynamics of this model is beyond
the scope of the present work and we leave these steps to future research. 

\section{Outlook}

The levitated helium drop offers a large number of unusual features
that represent opportunities for unconventional optomechanics and
fundamental studies of superfluid helium physics. Here we will briefly
mention some of those.

Due to the large energy of electronic transitions in helium, the drop
is expected to handle high circulating optical powers. We estimate
the optical spring effect in the drop to be $\sim1\,{\rm Hz}$ per
photon, so it should be possible to increase the drop's mechanical
frequencies by several orders of magnitude. It would remain to be
seen how the mechanical $Q$ of a given surface mode would evolve
as its frequency is increased past a large number of other mechanical
modes. At the same time, the static deflection can remain small ($\sim1\ \mu{\rm m}$)
even for $10^{8}$ circulating photons. Moreover, it would be rather
easy in this setup to reach the strong-coupling regime of linearized
optomechanics, $g_{0}N_{{\rm phot}}>\kappa$, for $g_{0}\sim{\rm 200}$
Hz and a conservative estimate of $\kappa\sim{\rm 10}$ kHz. Thus,
using the tools of linear optomechanics \cite{aspelmeyer_cavity_2014},
one could e.g. transfer nonclassical optical states into the surface
vibrational modes. Possibly, these could then be further transferred
onto the angular momentum state, generating novel optorotational control.
Alternatively, the dispersive measurement of angular momentum outlined
above can be used to generate interesting post-selected states, including
states of squeezed angular momentum.

Beyond the conventional linear optomechanical coupling, it should
also be possible to realize quadratic coupling in this setup. Indeed,
according to Eq.~(\ref{eq:GeneralCouplingConstant}), an optical
whispering-gallery mode whose plane is tilted at a particular angle
will have vanishing linear coupling to the equatorial bulge mode ($l=2$,
$m=0$), while the optical WGM in the equatorial plane itself has
no linear coupling to the $m\neq0$ mechanical modes.

In the present manuscript, we have entirely focussed on the lowest-order
surface vibration modes at $l=2$. However, one can imagine generating
interesting multimode optomechanics when addressing the higher-order
modes as well. The collective optical spring effect will be able to
generate an effective light-induced interaction between those modes,
which can get so strong as to form completely new normal modes. Moreover,
one can imagine exploiting transitions between optical modes of different
radial and angular momentum quantum numbers. These transitions will
then couple efficiently to higher $l$ mechanical modes, e.g. acoustic
whispering gallery modes, leading to Brillouin-like optomechanical
interactions \cite{bahl_observation_2012}.

When a drop's surface deformations or rotation rate become sufficiently
large, a variety of nonlinear effects are expected to occur. It is
known that a rotating drop can develop symmetry-broken shapes \cite{hill_nonaxisymmetric_2008},
but many questions remain open. For example, is it possible to obtain
stable drops with non-zero topological genus \cite{heine_computations_2006}?

Finally, the optical control and readout can serve as a completely
novel means to study the physics of superfluid helium in a setting
that is devoid of any complications arising from solid surfaces. For
example, at low temperatures, the damping of surface waves (ripplons)
is due to ripplon-phonon scattering. However, due to the finite size
of the drop, the bulk phonons inside the drop constitute a bath with
a very strongly frequency-dependent force noise spectrum and strongly
non-Markovian properties. These might be studied quantitatively, especially
using the optical spring effect as a tool to vary the ripplons' frequency.

Rotation in the superfluid drop is quantized and vortex lines emerge
as the drop is made to spin above a certain rotation rate \cite{seidel_morphology_1994,bauer_vortex_1995}.
Below that rate, the drop's angular momentum must be contained either
in surface modes or in the normal fluid (phonons propagating in the
bulk). The presence and the motion of the vortex lines then affects
the surface deformation, and this will be readily measurable optically.
Even a single vortex line is not inert. It can wiggle, and these vibrations
of the string-like vortex (known as Kelvin modes) could also be read
out via their effect on the optical WGM, providing a means for measuring
the mechanical properties of an isolated vortex line \cite{vinen_kelvin-wave_2003,kozik_kelvin-wave_2004}.
Moreover, one could investigate the interactions of many vortices
as well as quenches through phase transitions, e.g. observing Kibble-Zurek
type physics upon cooling a spinning drop. In general, optomechanics
in levitated helium drops may become a new tool enabling us to explore
a whole range of physical phenomena that are analogues to effects
in high-energy physics and cosmology \cite{volovik_universe_2009}.

\section*{Acknowledgments}

F.~M., M.S., and A.~A. acknowledge support through an ERC Starting
Grant (``OPTOMECH''), as well as the European FET proactive network
``Hybrid Optomechanical Technologies''.\textcolor{black}{{} L.C. acknowledges
support from a L'Oreal USA FWIS Fellowship (2012), NSERC Discovery
435554-2013, and a Canada Research Chairs grant 950-229003. J. H acknowledges
support from W. M. Keck Foundation Grant No. DT121914, AFOSR Grants
FA9550-09-1-0484 and FA9550-15-1-0270, DARPA Grant W911NF-14-1-0354,
ARO Grant W911NF-13-1-0104, and NSF Grant 1205861. This work has been
supported by the DARPA/MTO ORCHID Program through a grant from AFOSR.
This project was made possible through the support of a grant from
the John Templeton Foundation. The opinions expressed in this publication
are those of the authors and do not necessarily reflect the views
of the John Templeton Foundation. This material is based upon work
supported by the National Science Foundation Graduate Research Fellowship
under Grant No. DGE-1122492. }

\section*{Appendix A: Noise sources for the QND measurement of angular momentum}

There are three noise sources that may potentially interfere with
the QND measurement of the drop's angular momentum. 

The first is due to the fact that the evaporation of atoms is a stochastic
process. When $N$ atoms evaporate on average during a given time
interval, that number actually fluctuates by $\sqrt{N}$, leading
to corresponding fluctuations in the drop radius and the optical resonance.
The effect diminishes as the temperature decreases and the evaporation
rate slows. The relevant rates can be extracted from Fig.~\ref{Fig_EvaporativeCooling}.
After 1000 s of evaporation, a $^{3}\mathrm{{He}}$ drop reaches $T\approx0.13$
K with $\sim1$ nW of cooling power. This corresponds to $4\cdot10^{13}$
atoms evaporating per second, with a resulting deterministic drift
of the optical resonance of $\sim60$ MHz/s. In Section III.A we considered
a measurement time of 0.1 $\mu{\rm s}$, which is sufficiently long
to resolve an angular momentum spread of the order of the Heisenberg
uncertainty, $\sqrt{\hbar L}$. During this time, the number of evaporated
atoms fluctuates only by about $10^{3}$, leading to a negligible
stochastic optical shift of $\sim10^{-3}$ Hz. 

The second noise source is directly connected to the same physics:
the evaporating atoms will also carry away angular momentum. For $T=0.1\ {\rm K}$,
a single atom flying off with the mean thermal velocity can extract
$\sim10^{6}\hbar$ from a droplet of radius $R=1$ mm. Staying with
the example considered in the previous paragraph, in $0.1\ \mu\mathrm{s}$
this results in a stochastic contribution to $L_{z}$ of $10^{9}\hbar$,
much smaller than the $10^{11}\hbar$ measurement resolution mentioned
above.

Finally, the third noise source is present even in the absence of
evaporation. It consists of changes in the drop's angular momentum
due to the scattering of photons. Each randomly scattered photon can
carry away angular momentum $\sim R\cdot\hbar k$, which amounts to
about 6000 $\hbar$. Assuming an input power of 10 $\mu\mathrm{W}$
and that 10\% of the photons are scattered stochastically in random
directions (e.g. from the thermal surface fluctuations), this process
would result in a stochastic angular momentum transfer (during a 0.1
$\mu\mathrm{s}$ measurement time) of $\sim4\cdot10^{6}\hbar$, well
below the measurement uncertainty. 

\section*{Appendix B: Lagrangian for the coupling of rotations to the $l=2$
vibrations in an incompressible fluid drop}

The purpose of this appendix is to display the full Lagrangian describing
the coupling between arbitrary rotations and the vibrational $l=2$
surface modes of the drop. To that end, we have to introduce a number
of definitions. The derivation of this Lagrangian will be discussed
in a separate publication (see also the thesis \cite{schmidt_optomechanical_2015}).

For brevity, it is convenient from now on to measure lengths in units
of the sphere radius (such that $R=1$). Appropriate dimensions can
be re-instated later, if needed. The surface deformation pattern in
the laboratory frame is given by

\begin{equation}
\delta R^{{\rm Lab}}(\mathbf{r},t)=\sum_{m=-2}^{2}X_{m}^{{\rm Lab}}(t)\phi_{m}(\mathbf{r})\,,
\end{equation}
where $\mathbf{r}$ resides on the surface ($\left|\mathbf{r}\right|=1$).
The $\phi_{m}(\mathbf{r})$ are based on the $l=2$ spherical harmonics,
$\phi_{m}(\mathbf{r})\sim r^{2}Y_{l,m}(\theta,\phi)$. They have been
extended to cover all of space, which will simplify the notation further
below. More precisely, we have defined $\phi_{\pm2}=\mathcal{N}_{2}(x\pm iy)^{2}$,
$\phi_{\pm1}=\mathcal{N}_{1}(x\pm iy)z$, and $\phi_{0}=\mathcal{N}_{0}(x^{2}+y^{2}-2z^{2})$;
where the constants are $\mathcal{N}_{2}=(32\pi/15)^{-1/2}$, $\mathcal{N}_{1}=(8\pi/15)^{-1/2}$,
and $\mathcal{N}_{0}=(16\pi/5)^{-1/2}$. The surface integrals are
normalized, $\int\left|\phi_{m}\right|^{2}\sin\theta d\theta d\varphi=1$
for $|{\bf {r}}|$= 1.

To write down the Lagrangian, we need to convert between the lab frame
and the co-rotating frame (described by a set of three Euler angles
which we sometimes combine into a three-vector $\vec{\varphi}$).
We assume that the transformation is effected by a suitable $5\times5$
matrix $W$, with $X^{{\rm Lab}}=WX^{{\rm Rot}}$, or explicitly:

\begin{equation}
X_{m}^{{\rm Lab}}=\sum_{m=-2}^{2}W_{mm'}(\vec{\varphi})X_{m}^{{\rm Rot}}\,.
\end{equation}
Upon rotation of the drop by the angular frequency vector $\mathbf{\Omega}$
(which is expressed in the lab frame), the matrix $W$ changes according
to

\begin{equation}
\frac{d}{dt}W_{mm'}=-\sum_{s=1}^{3}\sum_{k=-2}^{2}\Omega_{s}K_{km}^{(s)}W_{km'}\,,
\end{equation}
or $\dot{W}=-\sum_{s}\Omega_{s}\left(K^{(s)}\right)^{t}W$ in matrix
notation. This relation defines the generators $K_{km}^{(s)}$ that
describe infinitesimal rotations. The generator $K^{(3)}$ for rotations
around the $z$-axis is the simplest one, with $K_{km}^{(3)}=im\delta_{k,m}$.
Finally, we introduce the notation $D_{m}^{{\rm Rot}}=\dot{X}_{m}^{{\rm Rot}}$,
and $D^{{\rm Lab}}=WD^{{\rm Rot}}$. With these definitions, we are
now in a position to write down the full Lagrangian that couples vibrations
and rotation:

\begin{eqnarray}
\mathcal{L} & = & \frac{I}{2}\mathbf{\Omega}^{2}+\frac{\rho}{4}\dot{X}_{m}^{{\rm Rot}*}\dot{X}_{m}^{{\rm Rot}}-\frac{I}{2}\delta R^{\mathrm{{Lab}}}(\mathbf{\Omega})\nonumber \\
 &  & +\frac{\rho}{4}D_{m}^{{\rm Lab}}\Omega_{s}K_{mm'}^{(s)}X_{m'}^{{\rm Lab}*}-2\sigma X_{m}^{{\rm Rot}*}X_{m}^{{\rm Rot}}.
\end{eqnarray}
Summation over repeated indices is implied. This Lagrangian contains,
in this order: (i) the rotational energy of the unperturbed spherical
drop, (ii) the kinetic energy of the surface vibrations, (iii) the
change in the rotational energy due to the deformation (with the surface
deformation field $\delta R$ evaluated at the angular momentum vector),
(iv) the term describing Bryan's effect (from the Coriolis force),
(v) the potential energy of the surface vibrations (due to the surface
tension). We note that all the deformation-related quantities ($X^{{\rm Rot}}$,
$X^{{\rm Lab}}$, and $D^{{\rm Lab}}$) have to be expressed via $X^{{\rm Rot}}$
for the purpose of deriving the equations of motion. We also note
that the $X^{{\rm Rot}}$ coefficients obey the constraint $X_{-m}^{{\rm Rot}}=X_{m}^{{\rm Rot}*}$
due to the fact that the surface deformation is real-valued. In deriving
the equations of motion, one can either split $X_{m}^{{\rm Rot}}$
into real and imaginary parts (for $m>0$) or, more efficiently, formally
treat $X_{m}^{{\rm Rot}}$ and $X_{m}^{{\rm Rot}*}$ as independent
variables.

\bibliographystyle{apsrev4-1}
\bibliography{HeliumDrop}

\begin{thebibliography}{78}%
\makeatletter
\providecommand \@ifxundefined [1]{%
 \@ifx{#1\undefined}
}%
\providecommand \@ifnum [1]{%
 \ifnum #1\expandafter \@firstoftwo
 \else \expandafter \@secondoftwo
 \fi
}%
\providecommand \@ifx [1]{%
 \ifx #1\expandafter \@firstoftwo
 \else \expandafter \@secondoftwo
 \fi
}%
\providecommand \natexlab [1]{#1}%
\providecommand \enquote  [1]{``#1''}%
\providecommand \bibnamefont  [1]{#1}%
\providecommand \bibfnamefont [1]{#1}%
\providecommand \citenamefont [1]{#1}%
\providecommand \href@noop [0]{\@secondoftwo}%
\providecommand \href [0]{\begingroup \@sanitize@url \@href}%
\providecommand \@href[1]{\@@startlink{#1}\@@href}%
\providecommand \@@href[1]{\endgroup#1\@@endlink}%
\providecommand \@sanitize@url [0]{\catcode `\\12\catcode `\$12\catcode
  `\&12\catcode `\#12\catcode `\^12\catcode `\_12\catcode `\%12\relax}%
\providecommand \@@startlink[1]{}%
\providecommand \@@endlink[0]{}%
\providecommand \url  [0]{\begingroup\@sanitize@url \@url }%
\providecommand \@url [1]{\endgroup\@href {#1}{\urlprefix }}%
\providecommand \urlprefix  [0]{URL }%
\providecommand \Eprint [0]{\href }%
\providecommand \doibase [0]{http://dx.doi.org/}%
\providecommand \selectlanguage [0]{\@gobble}%
\providecommand \bibinfo  [0]{\@secondoftwo}%
\providecommand \bibfield  [0]{\@secondoftwo}%
\providecommand \translation [1]{[#1]}%
\providecommand \BibitemOpen [0]{}%
\providecommand \bibitemStop [0]{}%
\providecommand \bibitemNoStop [0]{.\EOS\space}%
\providecommand \EOS [0]{\spacefactor3000\relax}%
\providecommand \BibitemShut  [1]{\csname bibitem#1\endcsname}%
\let\auto@bib@innerbib\@empty
\bibitem [{\citenamefont {Aspelmeyer}\ \emph {et~al.}(2014)\citenamefont
  {Aspelmeyer}, \citenamefont {Kippenberg},\ and\ \citenamefont
  {Marquardt}}]{aspelmeyer_cavity_2014}%
  \BibitemOpen
  \bibfield  {author} {\bibinfo {author} {\bibfnamefont {M.}~\bibnamefont
  {Aspelmeyer}}, \bibinfo {author} {\bibfnamefont {T.~J.}\ \bibnamefont
  {Kippenberg}}, \ and\ \bibinfo {author} {\bibfnamefont {F.}~\bibnamefont
  {Marquardt}},\ }\href {\doibase 10.1103/RevModPhys.86.1391} {\bibfield
  {journal} {\bibinfo  {journal} {Reviews of Modern Physics}\ }\textbf
  {\bibinfo {volume} {86}},\ \bibinfo {pages} {1391} (\bibinfo {year}
  {2014})}\BibitemShut {NoStop}%
\bibitem [{\citenamefont {O'Connell}\ \emph {et~al.}(2010)\citenamefont
  {O'Connell}, \citenamefont {Hofheinz}, \citenamefont {Ansmann}, \citenamefont
  {Bialczak}, \citenamefont {Lenander}, \citenamefont {Lucero}, \citenamefont
  {Neely}, \citenamefont {Sank}, \citenamefont {Wang}, \citenamefont {Weides},
  \citenamefont {Cleland},\ and\ \citenamefont
  {Martinis}}]{oconnell_quantum_2010}%
  \BibitemOpen
  \bibfield  {author} {\bibinfo {author} {\bibfnamefont {A.~D.}\ \bibnamefont
  {O'Connell}}, \bibinfo {author} {\bibfnamefont {M.}~\bibnamefont {Hofheinz}},
  \bibinfo {author} {\bibfnamefont {M.}~\bibnamefont {Ansmann}}, \bibinfo
  {author} {\bibfnamefont {R.~C.}\ \bibnamefont {Bialczak}}, \bibinfo {author}
  {\bibfnamefont {M.}~\bibnamefont {Lenander}}, \bibinfo {author}
  {\bibfnamefont {E.}~\bibnamefont {Lucero}}, \bibinfo {author} {\bibfnamefont
  {M.}~\bibnamefont {Neely}}, \bibinfo {author} {\bibfnamefont
  {D.}~\bibnamefont {Sank}}, \bibinfo {author} {\bibfnamefont {H.}~\bibnamefont
  {Wang}}, \bibinfo {author} {\bibfnamefont {M.}~\bibnamefont {Weides}},
  \bibinfo {author} {\bibfnamefont {A.~N.}\ \bibnamefont {Cleland}}, \ and\
  \bibinfo {author} {\bibfnamefont {J.~M.}\ \bibnamefont {Martinis}},\
  }\href@noop {} {\bibfield  {journal} {\bibinfo  {journal} {Nature}\ }\textbf
  {\bibinfo {volume} {464}},\ \bibinfo {pages} {697} (\bibinfo {year}
  {2010})}\BibitemShut {NoStop}%
\bibitem [{\citenamefont {Brahms}\ \emph {et~al.}(2012)\citenamefont {Brahms},
  \citenamefont {Botter}, \citenamefont {Schreppler}, \citenamefont {Brooks},\
  and\ \citenamefont {Stamper-Kurn}}]{brahms_optically_2012}%
  \BibitemOpen
  \bibfield  {author} {\bibinfo {author} {\bibfnamefont {N.}~\bibnamefont
  {Brahms}}, \bibinfo {author} {\bibfnamefont {T.}~\bibnamefont {Botter}},
  \bibinfo {author} {\bibfnamefont {S.}~\bibnamefont {Schreppler}}, \bibinfo
  {author} {\bibfnamefont {D.~W.~C.}\ \bibnamefont {Brooks}}, \ and\ \bibinfo
  {author} {\bibfnamefont {D.~M.}\ \bibnamefont {Stamper-Kurn}},\ }\href@noop
  {} {\bibfield  {journal} {\bibinfo  {journal} {Physical Review Letters}\
  }\textbf {\bibinfo {volume} {108}},\ \bibinfo {pages} {133601} (\bibinfo
  {year} {2012})}\BibitemShut {NoStop}%
\bibitem [{\citenamefont {Shkarin}\ \emph {et~al.}(2017)\citenamefont
  {Shkarin}, \citenamefont {Kashkanova}, \citenamefont {Brown}, \citenamefont
  {Hohmann}, \citenamefont {Ott}, \citenamefont {Reichel},\ and\ \citenamefont
  {Harris}}]{kashkanova_quantum_2017}%
  \BibitemOpen
  \bibfield  {author} {\bibinfo {author} {\bibfnamefont {A.~B.}\ \bibnamefont
  {Shkarin}}, \bibinfo {author} {\bibfnamefont {A.~D.}\ \bibnamefont
  {Kashkanova}}, \bibinfo {author} {\bibfnamefont {C.~D.}\ \bibnamefont
  {Brown}}, \bibinfo {author} {\bibfnamefont {L.}~\bibnamefont {Hohmann}},
  \bibinfo {author} {\bibfnamefont {K.}~\bibnamefont {Ott}}, \bibinfo {author}
  {\bibfnamefont {J.}~\bibnamefont {Reichel}}, \ and\ \bibinfo {author}
  {\bibfnamefont {J.~G.~E.}\ \bibnamefont {Harris}},\ }\href@noop {} {\bibfield
   {journal} {\bibinfo  {journal} {manuscript in preparation}\ } (\bibinfo
  {year} {2017})}\BibitemShut {NoStop}%
\bibitem [{\citenamefont {Purdy}\ \emph {et~al.}(2017)\citenamefont {Purdy},
  \citenamefont {Grutter}, \citenamefont {Srinivasan},\ and\ \citenamefont
  {Taylor}}]{purdy_quantum_2017}%
  \BibitemOpen
  \bibfield  {author} {\bibinfo {author} {\bibfnamefont {T.~P.}\ \bibnamefont
  {Purdy}}, \bibinfo {author} {\bibfnamefont {K.~E.}\ \bibnamefont {Grutter}},
  \bibinfo {author} {\bibfnamefont {K.}~\bibnamefont {Srinivasan}}, \ and\
  \bibinfo {author} {\bibfnamefont {J.~M.}\ \bibnamefont {Taylor}},\ }\href
  {\doibase 10.1126/science.aag1407} {\bibfield  {journal} {\bibinfo  {journal}
  {Science}\ }\textbf {\bibinfo {volume} {356}},\ \bibinfo {pages} {1265}
  (\bibinfo {year} {2017})}\BibitemShut {NoStop}%
\bibitem [{\citenamefont {Underwood}\ \emph {et~al.}(2015)\citenamefont
  {Underwood}, \citenamefont {Mason}, \citenamefont {Lee}, \citenamefont {Xu},
  \citenamefont {Jiang}, \citenamefont {Shkarin}, \citenamefont {B{\o}rkje},
  \citenamefont {Girvin},\ and\ \citenamefont
  {Harris}}]{underwood_measurement_2015}%
  \BibitemOpen
  \bibfield  {author} {\bibinfo {author} {\bibfnamefont {M.}~\bibnamefont
  {Underwood}}, \bibinfo {author} {\bibfnamefont {D.}~\bibnamefont {Mason}},
  \bibinfo {author} {\bibfnamefont {D.}~\bibnamefont {Lee}}, \bibinfo {author}
  {\bibfnamefont {H.}~\bibnamefont {Xu}}, \bibinfo {author} {\bibfnamefont
  {L.}~\bibnamefont {Jiang}}, \bibinfo {author} {\bibfnamefont {A.~B.}\
  \bibnamefont {Shkarin}}, \bibinfo {author} {\bibfnamefont {K.}~\bibnamefont
  {B{\o}rkje}}, \bibinfo {author} {\bibfnamefont {S.~M.}\ \bibnamefont
  {Girvin}}, \ and\ \bibinfo {author} {\bibfnamefont {J.~G.~E.}\ \bibnamefont
  {Harris}},\ }\href {\doibase 10.1103/PhysRevA.92.061801} {\bibfield
  {journal} {\bibinfo  {journal} {Physical Review A}\ }\textbf {\bibinfo
  {volume} {92}},\ \bibinfo {pages} {061801} (\bibinfo {year}
  {2015})}\BibitemShut {NoStop}%
\bibitem [{\citenamefont {Pikovski}\ \emph {et~al.}(2012)\citenamefont
  {Pikovski}, \citenamefont {Vanner}, \citenamefont {Aspelmeyer}, \citenamefont
  {Kim},\ and\ \citenamefont {Brukner}}]{pikovski_probing_2012}%
  \BibitemOpen
  \bibfield  {author} {\bibinfo {author} {\bibfnamefont {I.}~\bibnamefont
  {Pikovski}}, \bibinfo {author} {\bibfnamefont {M.~R.}\ \bibnamefont
  {Vanner}}, \bibinfo {author} {\bibfnamefont {M.}~\bibnamefont {Aspelmeyer}},
  \bibinfo {author} {\bibfnamefont {M.~S.}\ \bibnamefont {Kim}}, \ and\
  \bibinfo {author} {\bibfnamefont {C.}~\bibnamefont {Brukner}},\ }\href
  {\doibase 10.1038/NPHYS2262} {\bibfield  {journal} {\bibinfo  {journal}
  {Nature Physics}\ }\textbf {\bibinfo {volume} {8}},\ \bibinfo {pages} {393}
  (\bibinfo {year} {2012})}\BibitemShut {NoStop}%
\bibitem [{\citenamefont {Chang}\ \emph {et~al.}(2010)\citenamefont {Chang},
  \citenamefont {Regal}, \citenamefont {Papp}, \citenamefont {Wilson},
  \citenamefont {Ye}, \citenamefont {Painter}, \citenamefont {Kimble},\ and\
  \citenamefont {Zoller}}]{chang_cavity_2010}%
  \BibitemOpen
  \bibfield  {author} {\bibinfo {author} {\bibfnamefont {D.~E.}\ \bibnamefont
  {Chang}}, \bibinfo {author} {\bibfnamefont {C.~A.}\ \bibnamefont {Regal}},
  \bibinfo {author} {\bibfnamefont {S.~B.}\ \bibnamefont {Papp}}, \bibinfo
  {author} {\bibfnamefont {D.~J.}\ \bibnamefont {Wilson}}, \bibinfo {author}
  {\bibfnamefont {J.}~\bibnamefont {Ye}}, \bibinfo {author} {\bibfnamefont
  {O.}~\bibnamefont {Painter}}, \bibinfo {author} {\bibfnamefont {H.~J.}\
  \bibnamefont {Kimble}}, \ and\ \bibinfo {author} {\bibfnamefont
  {P.}~\bibnamefont {Zoller}},\ }\href {\doibase 10.1073/pnas.0912969107}
  {\bibfield  {journal} {\bibinfo  {journal} {Proceedings of the National
  Academy of Sciences}\ }\textbf {\bibinfo {volume} {107}},\ \bibinfo {pages}
  {1005} (\bibinfo {year} {2010})}\BibitemShut {NoStop}%
\bibitem [{\citenamefont {Romero-Isart}\ \emph {et~al.}(2010)\citenamefont
  {Romero-Isart}, \citenamefont {Juan}, \citenamefont {Quidant},\ and\
  \citenamefont {Cirac}}]{romero-isart_toward_2010}%
  \BibitemOpen
  \bibfield  {author} {\bibinfo {author} {\bibfnamefont {O.}~\bibnamefont
  {Romero-Isart}}, \bibinfo {author} {\bibfnamefont {M.~L.}\ \bibnamefont
  {Juan}}, \bibinfo {author} {\bibfnamefont {R.}~\bibnamefont {Quidant}}, \
  and\ \bibinfo {author} {\bibfnamefont {J.~I.}\ \bibnamefont {Cirac}},\ }\href
  {http://iopscience.iop.org/1367-2630/12/3/033015/video/abstract} {\bibfield
  {journal} {\bibinfo  {journal} {New Journal of Physics}\ }\textbf {\bibinfo
  {volume} {12}},\ \bibinfo {pages} {033015} (\bibinfo {year}
  {2010})}\BibitemShut {NoStop}%
\bibitem [{\citenamefont {Barker}\ and\ \citenamefont
  {Shneider}(2010)}]{barker_cavity_2010}%
  \BibitemOpen
  \bibfield  {author} {\bibinfo {author} {\bibfnamefont {P.~F.}\ \bibnamefont
  {Barker}}\ and\ \bibinfo {author} {\bibfnamefont {M.~N.}\ \bibnamefont
  {Shneider}},\ }\href {\doibase 10.1103/PhysRevA.81.023826} {\bibfield
  {journal} {\bibinfo  {journal} {Physical Review A}\ }\textbf {\bibinfo
  {volume} {81}},\ \bibinfo {pages} {023826} (\bibinfo {year}
  {2010})}\BibitemShut {NoStop}%
\bibitem [{\citenamefont {Millen}\ \emph {et~al.}(2014)\citenamefont {Millen},
  \citenamefont {Deesuwan}, \citenamefont {Barker},\ and\ \citenamefont
  {Anders}}]{millen_nanoscale_2014}%
  \BibitemOpen
  \bibfield  {author} {\bibinfo {author} {\bibfnamefont {J.}~\bibnamefont
  {Millen}}, \bibinfo {author} {\bibfnamefont {T.}~\bibnamefont {Deesuwan}},
  \bibinfo {author} {\bibfnamefont {P.}~\bibnamefont {Barker}}, \ and\ \bibinfo
  {author} {\bibfnamefont {J.}~\bibnamefont {Anders}},\ }\href {\doibase
  10.1038/nnano.2014.82} {\bibfield  {journal} {\bibinfo  {journal} {Nature
  Nanotechnology}\ }\textbf {\bibinfo {volume} {9}},\ \bibinfo {pages} {425}
  (\bibinfo {year} {2014})}\BibitemShut {NoStop}%
\bibitem [{\citenamefont {Li}\ \emph {et~al.}(2010)\citenamefont {Li},
  \citenamefont {Kheifets}, \citenamefont {Medellin},\ and\ \citenamefont
  {Raizen}}]{li_measurement_2010}%
  \BibitemOpen
  \bibfield  {author} {\bibinfo {author} {\bibfnamefont {T.}~\bibnamefont
  {Li}}, \bibinfo {author} {\bibfnamefont {S.}~\bibnamefont {Kheifets}},
  \bibinfo {author} {\bibfnamefont {D.}~\bibnamefont {Medellin}}, \ and\
  \bibinfo {author} {\bibfnamefont {M.~G.}\ \bibnamefont {Raizen}},\ }\href
  {\doibase 10.1126/science.1189403} {\bibfield  {journal} {\bibinfo  {journal}
  {Science}\ }\textbf {\bibinfo {volume} {328}},\ \bibinfo {pages} {1673}
  (\bibinfo {year} {2010})}\BibitemShut {NoStop}%
\bibitem [{\citenamefont {Li}\ \emph {et~al.}(2011)\citenamefont {Li},
  \citenamefont {Kheifets},\ and\ \citenamefont
  {Raizen}}]{li_millikelvin_2011}%
  \BibitemOpen
  \bibfield  {author} {\bibinfo {author} {\bibfnamefont {T.}~\bibnamefont
  {Li}}, \bibinfo {author} {\bibfnamefont {S.}~\bibnamefont {Kheifets}}, \ and\
  \bibinfo {author} {\bibfnamefont {M.~G.}\ \bibnamefont {Raizen}},\ }\href
  {\doibase 10.1038/nphys1952} {\bibfield  {journal} {\bibinfo  {journal}
  {Nature Physics}\ }\textbf {\bibinfo {volume} {7}},\ \bibinfo {pages} {527}
  (\bibinfo {year} {2011})}\BibitemShut {NoStop}%
\bibitem [{\citenamefont {Romero-Isart}\ \emph {et~al.}(2011)\citenamefont
  {Romero-Isart}, \citenamefont {Pflanzer}, \citenamefont {Blaser},
  \citenamefont {Kaltenbaek}, \citenamefont {Kiesel}, \citenamefont
  {Aspelmeyer},\ and\ \citenamefont {Cirac}}]{romero-isart_large_2011}%
  \BibitemOpen
  \bibfield  {author} {\bibinfo {author} {\bibfnamefont {O.}~\bibnamefont
  {Romero-Isart}}, \bibinfo {author} {\bibfnamefont {A.~C.}\ \bibnamefont
  {Pflanzer}}, \bibinfo {author} {\bibfnamefont {F.}~\bibnamefont {Blaser}},
  \bibinfo {author} {\bibfnamefont {R.}~\bibnamefont {Kaltenbaek}}, \bibinfo
  {author} {\bibfnamefont {N.}~\bibnamefont {Kiesel}}, \bibinfo {author}
  {\bibfnamefont {M.}~\bibnamefont {Aspelmeyer}}, \ and\ \bibinfo {author}
  {\bibfnamefont {J.~I.}\ \bibnamefont {Cirac}},\ }\href {\doibase
  10.1103/PhysRevLett.107.020405} {\bibfield  {journal} {\bibinfo  {journal}
  {Physical Review Letters}\ }\textbf {\bibinfo {volume} {107}},\ \bibinfo
  {pages} {020405} (\bibinfo {year} {2011})}\BibitemShut {NoStop}%
\bibitem [{\citenamefont {Gieseler}\ \emph {et~al.}(2012)\citenamefont
  {Gieseler}, \citenamefont {Deutsch}, \citenamefont {Quidant},\ and\
  \citenamefont {Novotny}}]{gieseler_subkelvin_2012}%
  \BibitemOpen
  \bibfield  {author} {\bibinfo {author} {\bibfnamefont {J.}~\bibnamefont
  {Gieseler}}, \bibinfo {author} {\bibfnamefont {B.}~\bibnamefont {Deutsch}},
  \bibinfo {author} {\bibfnamefont {R.}~\bibnamefont {Quidant}}, \ and\
  \bibinfo {author} {\bibfnamefont {L.}~\bibnamefont {Novotny}},\ }\href
  {\doibase 10.1103/PhysRevLett.109.103603} {\bibfield  {journal} {\bibinfo
  {journal} {Physical Review Letters}\ }\textbf {\bibinfo {volume} {109}},\
  \bibinfo {pages} {103603} (\bibinfo {year} {2012})}\BibitemShut {NoStop}%
\bibitem [{\citenamefont {Murch}\ \emph {et~al.}(2008)\citenamefont {Murch},
  \citenamefont {Moore}, \citenamefont {Gupta},\ and\ \citenamefont
  {Stamper-Kurn}}]{murch_observation_2008}%
  \BibitemOpen
  \bibfield  {author} {\bibinfo {author} {\bibfnamefont {K.~W.}\ \bibnamefont
  {Murch}}, \bibinfo {author} {\bibfnamefont {K.~L.}\ \bibnamefont {Moore}},
  \bibinfo {author} {\bibfnamefont {S.}~\bibnamefont {Gupta}}, \ and\ \bibinfo
  {author} {\bibfnamefont {D.~M.}\ \bibnamefont {Stamper-Kurn}},\ }\href
  {\doibase 10.1038/nphys965} {\bibfield  {journal} {\bibinfo  {journal}
  {Nature Physics}\ }\textbf {\bibinfo {volume} {4}},\ \bibinfo {pages} {561}
  (\bibinfo {year} {2008})}\BibitemShut {NoStop}%
\bibitem [{\citenamefont {Brennecke}\ \emph {et~al.}(2008)\citenamefont
  {Brennecke}, \citenamefont {Ritter}, \citenamefont {Donner},\ and\
  \citenamefont {Esslinger}}]{brennecke_cavity_2008}%
  \BibitemOpen
  \bibfield  {author} {\bibinfo {author} {\bibfnamefont {F.}~\bibnamefont
  {Brennecke}}, \bibinfo {author} {\bibfnamefont {S.}~\bibnamefont {Ritter}},
  \bibinfo {author} {\bibfnamefont {T.}~\bibnamefont {Donner}}, \ and\ \bibinfo
  {author} {\bibfnamefont {T.}~\bibnamefont {Esslinger}},\ }\href {\doibase
  10.1126/science.1163218} {\bibfield  {journal} {\bibinfo  {journal}
  {Science}\ }\textbf {\bibinfo {volume} {322}},\ \bibinfo {pages} {235}
  (\bibinfo {year} {2008})}\BibitemShut {NoStop}%
\bibitem [{\citenamefont {Dahan}\ \emph {et~al.}(2016)\citenamefont {Dahan},
  \citenamefont {Martin},\ and\ \citenamefont {Carmon}}]{dahan_droplet_2016}%
  \BibitemOpen
  \bibfield  {author} {\bibinfo {author} {\bibfnamefont {R.}~\bibnamefont
  {Dahan}}, \bibinfo {author} {\bibfnamefont {L.~L.}\ \bibnamefont {Martin}}, \
  and\ \bibinfo {author} {\bibfnamefont {T.}~\bibnamefont {Carmon}},\ }\href
  {\doibase 10.1364/OPTICA.3.000175} {\bibfield  {journal} {\bibinfo  {journal}
  {Optica}\ }\textbf {\bibinfo {volume} {3}},\ \bibinfo {pages} {175} (\bibinfo
  {year} {2016})}\BibitemShut {NoStop}%
\bibitem [{\citenamefont {Bahl}\ \emph {et~al.}(2013)\citenamefont {Bahl},
  \citenamefont {Kim}, \citenamefont {Lee}, \citenamefont {Liu}, \citenamefont
  {Fan},\ and\ \citenamefont {Carmon}}]{bahl_brillouin_2013}%
  \BibitemOpen
  \bibfield  {author} {\bibinfo {author} {\bibfnamefont {G.}~\bibnamefont
  {Bahl}}, \bibinfo {author} {\bibfnamefont {K.~H.}\ \bibnamefont {Kim}},
  \bibinfo {author} {\bibfnamefont {W.}~\bibnamefont {Lee}}, \bibinfo {author}
  {\bibfnamefont {J.}~\bibnamefont {Liu}}, \bibinfo {author} {\bibfnamefont
  {X.}~\bibnamefont {Fan}}, \ and\ \bibinfo {author} {\bibfnamefont
  {T.}~\bibnamefont {Carmon}},\ }\href {\doibase 10.1038/ncomms2994} {\bibfield
   {journal} {\bibinfo  {journal} {Nature Communications}\ }\textbf {\bibinfo
  {volume} {4}},\ \bibinfo {pages} {1994} (\bibinfo {year} {2013})}\BibitemShut
  {NoStop}%
\bibitem [{\citenamefont {Kashkanova}\ \emph
  {et~al.}(2017{\natexlab{a}})\citenamefont {Kashkanova}, \citenamefont
  {Shkarin}, \citenamefont {Brown}, \citenamefont {Flowers-Jacobs},
  \citenamefont {Childress}, \citenamefont {Hoch}, \citenamefont {Hohmann},
  \citenamefont {Ott}, \citenamefont {Reichel},\ and\ \citenamefont
  {Harris}}]{kashkanova_superfluid_2017}%
  \BibitemOpen
  \bibfield  {author} {\bibinfo {author} {\bibfnamefont {A.~D.}\ \bibnamefont
  {Kashkanova}}, \bibinfo {author} {\bibfnamefont {A.~B.}\ \bibnamefont
  {Shkarin}}, \bibinfo {author} {\bibfnamefont {C.~D.}\ \bibnamefont {Brown}},
  \bibinfo {author} {\bibfnamefont {N.~E.}\ \bibnamefont {Flowers-Jacobs}},
  \bibinfo {author} {\bibfnamefont {L.}~\bibnamefont {Childress}}, \bibinfo
  {author} {\bibfnamefont {S.~W.}\ \bibnamefont {Hoch}}, \bibinfo {author}
  {\bibfnamefont {L.}~\bibnamefont {Hohmann}}, \bibinfo {author} {\bibfnamefont
  {K.}~\bibnamefont {Ott}}, \bibinfo {author} {\bibfnamefont {J.}~\bibnamefont
  {Reichel}}, \ and\ \bibinfo {author} {\bibfnamefont {J.~G.~E.}\ \bibnamefont
  {Harris}},\ }\href {\doibase 10.1038/nphys3900} {\bibfield  {journal}
  {\bibinfo  {journal} {Nature Physics}\ }\textbf {\bibinfo {volume} {13}},\
  \bibinfo {pages} {74} (\bibinfo {year} {2017}{\natexlab{a}})}\BibitemShut
  {NoStop}%
\bibitem [{\citenamefont {Kashkanova}\ \emph
  {et~al.}(2017{\natexlab{b}})\citenamefont {Kashkanova}, \citenamefont
  {Shkarin}, \citenamefont {Brown}, \citenamefont {Flowers-Jacobs},
  \citenamefont {Childress}, \citenamefont {Hoch}, \citenamefont {Hohmann},
  \citenamefont {Ott}, \citenamefont {{J Reichel}},\ and\ \citenamefont
  {Harris}}]{kashkanova_optomechanics_2017}%
  \BibitemOpen
  \bibfield  {author} {\bibinfo {author} {\bibfnamefont {A.~D.}\ \bibnamefont
  {Kashkanova}}, \bibinfo {author} {\bibfnamefont {A.~B.}\ \bibnamefont
  {Shkarin}}, \bibinfo {author} {\bibfnamefont {C.~D.}\ \bibnamefont {Brown}},
  \bibinfo {author} {\bibfnamefont {N.~E.}\ \bibnamefont {Flowers-Jacobs}},
  \bibinfo {author} {\bibfnamefont {L.}~\bibnamefont {Childress}}, \bibinfo
  {author} {\bibfnamefont {S.~W.}\ \bibnamefont {Hoch}}, \bibinfo {author}
  {\bibfnamefont {L.}~\bibnamefont {Hohmann}}, \bibinfo {author} {\bibfnamefont
  {K.}~\bibnamefont {Ott}}, \bibinfo {author} {\bibnamefont {{J Reichel}}}, \
  and\ \bibinfo {author} {\bibfnamefont {J.~G.~E.}\ \bibnamefont {Harris}},\
  }\href {\doibase 10.1088/2040-8986/aa551e} {\bibfield  {journal} {\bibinfo
  {journal} {Journal of Optics}\ }\textbf {\bibinfo {volume} {19}},\ \bibinfo
  {pages} {034001} (\bibinfo {year} {2017}{\natexlab{b}})}\BibitemShut
  {NoStop}%
\bibitem [{\citenamefont {Lorenzo}\ and\ \citenamefont
  {Schwab}(2014)}]{lorenzo_superfluid_2014}%
  \BibitemOpen
  \bibfield  {author} {\bibinfo {author} {\bibfnamefont {L.~A.~D.}\
  \bibnamefont {Lorenzo}}\ and\ \bibinfo {author} {\bibfnamefont {K.~C.}\
  \bibnamefont {Schwab}},\ }\href {\doibase 10.1088/1367-2630/16/11/113020}
  {\bibfield  {journal} {\bibinfo  {journal} {New Journal of Physics}\ }\textbf
  {\bibinfo {volume} {16}},\ \bibinfo {pages} {113020} (\bibinfo {year}
  {2014})}\BibitemShut {NoStop}%
\bibitem [{\citenamefont {Harris}\ \emph {et~al.}(2016)\citenamefont {Harris},
  \citenamefont {McAuslan}, \citenamefont {Sheridan}, \citenamefont {Sachkou},
  \citenamefont {Baker},\ and\ \citenamefont {Bowen}}]{harris_laser_2016}%
  \BibitemOpen
  \bibfield  {author} {\bibinfo {author} {\bibfnamefont {G.~I.}\ \bibnamefont
  {Harris}}, \bibinfo {author} {\bibfnamefont {D.~L.}\ \bibnamefont
  {McAuslan}}, \bibinfo {author} {\bibfnamefont {E.}~\bibnamefont {Sheridan}},
  \bibinfo {author} {\bibfnamefont {Y.}~\bibnamefont {Sachkou}}, \bibinfo
  {author} {\bibfnamefont {C.}~\bibnamefont {Baker}}, \ and\ \bibinfo {author}
  {\bibfnamefont {W.~P.}\ \bibnamefont {Bowen}},\ }\href {\doibase
  10.1038/nphys3714} {\bibfield  {journal} {\bibinfo  {journal} {Nature
  Physics}\ }\textbf {\bibinfo {volume} {12}},\ \bibinfo {pages} {788}
  (\bibinfo {year} {2016})}\BibitemShut {NoStop}%
\bibitem [{\citenamefont {Singh}\ \emph {et~al.}(2016)\citenamefont {Singh},
  \citenamefont {De~Lorenzo}, \citenamefont {Pikovski},\ and\ \citenamefont
  {Schwab}}]{singh_detecting_2016}%
  \BibitemOpen
  \bibfield  {author} {\bibinfo {author} {\bibfnamefont {S.}~\bibnamefont
  {Singh}}, \bibinfo {author} {\bibfnamefont {L.~A.}\ \bibnamefont
  {De~Lorenzo}}, \bibinfo {author} {\bibfnamefont {I.}~\bibnamefont
  {Pikovski}}, \ and\ \bibinfo {author} {\bibfnamefont {K.~C.}\ \bibnamefont
  {Schwab}},\ }\href {http://arxiv.org/abs/1606.04980} {\bibfield  {journal}
  {\bibinfo  {journal} {arXiv:1606.04980 [astro-ph, physics:cond-mat,
  physics:gr-qc, physics:quant-ph]}\ } (\bibinfo {year} {2016})}\BibitemShut
  {NoStop}%
\bibitem [{gom()}]{gomez_shapes_2014}%
  \BibitemOpen
  \href@noop {} {}\bibinfo {note} {L. F. Gomez et al., Science {\bf 345}, 906
  (2014)}\BibitemShut {NoStop}%
\bibitem [{\citenamefont {Bauer}\ \emph {et~al.}(1995)\citenamefont {Bauer},
  \citenamefont {Donnelly},\ and\ \citenamefont {Vinen}}]{bauer_vortex_1995}%
  \BibitemOpen
  \bibfield  {author} {\bibinfo {author} {\bibfnamefont {G.~H.}\ \bibnamefont
  {Bauer}}, \bibinfo {author} {\bibfnamefont {R.~H.}\ \bibnamefont {Donnelly}},
  \ and\ \bibinfo {author} {\bibfnamefont {W.~F.}\ \bibnamefont {Vinen}},\
  }\href@noop {} {\bibfield  {journal} {\bibinfo  {journal} {Journal of Low
  Temerature Physics}\ }\textbf {\bibinfo {volume} {98}},\ \bibinfo {pages}
  {47} (\bibinfo {year} {1995})}\BibitemShut {NoStop}%
\bibitem [{\citenamefont {Weilert}\ \emph {et~al.}(1996)\citenamefont
  {Weilert}, \citenamefont {Whitaker}, \citenamefont {Maris},\ and\
  \citenamefont {Seidel}}]{weilert_magnetic_1996}%
  \BibitemOpen
  \bibfield  {author} {\bibinfo {author} {\bibfnamefont {M.~A.}\ \bibnamefont
  {Weilert}}, \bibinfo {author} {\bibfnamefont {D.~L.}\ \bibnamefont
  {Whitaker}}, \bibinfo {author} {\bibfnamefont {H.~J.}\ \bibnamefont {Maris}},
  \ and\ \bibinfo {author} {\bibfnamefont {G.~M.}\ \bibnamefont {Seidel}},\
  }\href {\doibase 10.1103/PhysRevLett.77.4840} {\bibfield  {journal} {\bibinfo
   {journal} {Physical Review Letters}\ }\textbf {\bibinfo {volume} {77}},\
  \bibinfo {pages} {4840} (\bibinfo {year} {1996})}\BibitemShut {NoStop}%
\bibitem [{\citenamefont {Weilert}\ \emph {et~al.}(1997)\citenamefont
  {Weilert}, \citenamefont {Whitaker}, \citenamefont {Maris},\ and\
  \citenamefont {Seidel}}]{weilert_magnetic_1997}%
  \BibitemOpen
  \bibfield  {author} {\bibinfo {author} {\bibfnamefont {M.~A.}\ \bibnamefont
  {Weilert}}, \bibinfo {author} {\bibfnamefont {D.~L.}\ \bibnamefont
  {Whitaker}}, \bibinfo {author} {\bibfnamefont {H.~J.}\ \bibnamefont {Maris}},
  \ and\ \bibinfo {author} {\bibfnamefont {G.~M.}\ \bibnamefont {Seidel}},\
  }\href {\doibase 10.1007/BF02403919} {\bibfield  {journal} {\bibinfo
  {journal} {Journal of Low Temperature Physics}\ }\textbf {\bibinfo {volume}
  {106}},\ \bibinfo {pages} {101} (\bibinfo {year} {1997})}\BibitemShut
  {NoStop}%
\bibitem [{\citenamefont {Vicente}\ \emph {et~al.}(2002)\citenamefont
  {Vicente}, \citenamefont {Yao}, \citenamefont {Maris},\ and\ \citenamefont
  {Seidel}}]{vicente_surface_2002}%
  \BibitemOpen
  \bibfield  {author} {\bibinfo {author} {\bibfnamefont {C.}~\bibnamefont
  {Vicente}}, \bibinfo {author} {\bibfnamefont {W.}~\bibnamefont {Yao}},
  \bibinfo {author} {\bibfnamefont {H.~J.}\ \bibnamefont {Maris}}, \ and\
  \bibinfo {author} {\bibfnamefont {G.~M.}\ \bibnamefont {Seidel}},\ }\href
  {\doibase 10.1103/PhysRevB.66.214504} {\bibfield  {journal} {\bibinfo
  {journal} {Physical Review B}\ }\textbf {\bibinfo {volume} {66}},\ \bibinfo
  {pages} {214504} (\bibinfo {year} {2002})}\BibitemShut {NoStop}%
\bibitem [{\citenamefont {Toennies}\ and\ \citenamefont
  {Vilesov}(2004)}]{toennies_superfluid_2004}%
  \BibitemOpen
  \bibfield  {author} {\bibinfo {author} {\bibfnamefont {J.~P.}\ \bibnamefont
  {Toennies}}\ and\ \bibinfo {author} {\bibfnamefont {A.~F.}\ \bibnamefont
  {Vilesov}},\ }\href {\doibase 10.1002/anie.200300611} {\bibfield  {journal}
  {\bibinfo  {journal} {Angewandte Chemie International Edition}\ }\textbf
  {\bibinfo {volume} {43}},\ \bibinfo {pages} {2622} (\bibinfo {year}
  {2004})}\BibitemShut {NoStop}%
\bibitem [{\citenamefont {Tzeng}\ \emph {et~al.}(1984)\citenamefont {Tzeng},
  \citenamefont {Wall}, \citenamefont {Long},\ and\ \citenamefont
  {Chang}}]{tzeng_laser_1984}%
  \BibitemOpen
  \bibfield  {author} {\bibinfo {author} {\bibfnamefont {H.-M.}\ \bibnamefont
  {Tzeng}}, \bibinfo {author} {\bibfnamefont {K.~F.}\ \bibnamefont {Wall}},
  \bibinfo {author} {\bibfnamefont {M.~B.}\ \bibnamefont {Long}}, \ and\
  \bibinfo {author} {\bibfnamefont {R.~K.}\ \bibnamefont {Chang}},\ }\href
  {\doibase 10.1364/OL.9.000499} {\bibfield  {journal} {\bibinfo  {journal}
  {Optics Letters}\ }\textbf {\bibinfo {volume} {9}},\ \bibinfo {pages} {499}
  (\bibinfo {year} {1984})}\BibitemShut {NoStop}%
\bibitem [{\citenamefont {Tzeng}\ \emph {et~al.}(1985)\citenamefont {Tzeng},
  \citenamefont {Barber}, \citenamefont {Long},\ and\ \citenamefont
  {Chang}}]{tzeng_laser-induced_1985}%
  \BibitemOpen
  \bibfield  {author} {\bibinfo {author} {\bibfnamefont {H.-M.}\ \bibnamefont
  {Tzeng}}, \bibinfo {author} {\bibfnamefont {P.~W.}\ \bibnamefont {Barber}},
  \bibinfo {author} {\bibfnamefont {M.~B.}\ \bibnamefont {Long}}, \ and\
  \bibinfo {author} {\bibfnamefont {R.~K.}\ \bibnamefont {Chang}},\ }\href
  {\doibase 10.1364/OL.10.000209} {\bibfield  {journal} {\bibinfo  {journal}
  {Optics Letters}\ }\textbf {\bibinfo {volume} {10}},\ \bibinfo {pages} {209}
  (\bibinfo {year} {1985})}\BibitemShut {NoStop}%
\bibitem [{\citenamefont {Qian}\ \emph {et~al.}(1986)\citenamefont {Qian},
  \citenamefont {Snow}, \citenamefont {Tzeng},\ and\ \citenamefont
  {Chang}}]{qian_lasing_1986}%
  \BibitemOpen
  \bibfield  {author} {\bibinfo {author} {\bibfnamefont {S.-X.}\ \bibnamefont
  {Qian}}, \bibinfo {author} {\bibfnamefont {J.~B.}\ \bibnamefont {Snow}},
  \bibinfo {author} {\bibfnamefont {H.-M.}\ \bibnamefont {Tzeng}}, \ and\
  \bibinfo {author} {\bibfnamefont {R.~K.}\ \bibnamefont {Chang}},\ }\href
  {\doibase 10.1126/science.231.4737.486} {\bibfield  {journal} {\bibinfo
  {journal} {Science}\ }\textbf {\bibinfo {volume} {231}},\ \bibinfo {pages}
  {486} (\bibinfo {year} {1986})}\BibitemShut {NoStop}%
\bibitem [{\citenamefont {Tanyeri}\ \emph {et~al.}(2007)\citenamefont
  {Tanyeri}, \citenamefont {Perron},\ and\ \citenamefont
  {Kennedy}}]{tanyeri_lasing_2007}%
  \BibitemOpen
  \bibfield  {author} {\bibinfo {author} {\bibfnamefont {M.}~\bibnamefont
  {Tanyeri}}, \bibinfo {author} {\bibfnamefont {R.}~\bibnamefont {Perron}}, \
  and\ \bibinfo {author} {\bibfnamefont {I.~M.}\ \bibnamefont {Kennedy}},\
  }\href {\doibase 10.1364/OL.32.002529} {\bibfield  {journal} {\bibinfo
  {journal} {Optics Letters}\ }\textbf {\bibinfo {volume} {32}},\ \bibinfo
  {pages} {2529} (\bibinfo {year} {2007})}\BibitemShut {NoStop}%
\bibitem [{\citenamefont {Uetake}\ \emph {et~al.}(1999)\citenamefont {Uetake},
  \citenamefont {Katsuragawa}, \citenamefont {Suzuki},\ and\ \citenamefont
  {Hakuta}}]{uetake_stimulated_1999}%
  \BibitemOpen
  \bibfield  {author} {\bibinfo {author} {\bibfnamefont {S.}~\bibnamefont
  {Uetake}}, \bibinfo {author} {\bibfnamefont {M.}~\bibnamefont {Katsuragawa}},
  \bibinfo {author} {\bibfnamefont {M.}~\bibnamefont {Suzuki}}, \ and\ \bibinfo
  {author} {\bibfnamefont {K.}~\bibnamefont {Hakuta}},\ }\href {\doibase
  10.1103/PhysRevA.61.011803} {\bibfield  {journal} {\bibinfo  {journal}
  {Physical Review A}\ }\textbf {\bibinfo {volume} {61}},\ \bibinfo {pages}
  {011803} (\bibinfo {year} {1999})}\BibitemShut {NoStop}%
\bibitem [{\citenamefont {Uetake}\ \emph {et~al.}(2002)\citenamefont {Uetake},
  \citenamefont {Sihombing},\ and\ \citenamefont
  {Hakuta}}]{uetake_stimulated_2002}%
  \BibitemOpen
  \bibfield  {author} {\bibinfo {author} {\bibfnamefont {S.}~\bibnamefont
  {Uetake}}, \bibinfo {author} {\bibfnamefont {R.~S.~D.}\ \bibnamefont
  {Sihombing}}, \ and\ \bibinfo {author} {\bibfnamefont {K.}~\bibnamefont
  {Hakuta}},\ }\href {\doibase 10.1364/OL.27.000421} {\bibfield  {journal}
  {\bibinfo  {journal} {Optics Letters}\ }\textbf {\bibinfo {volume} {27}},\
  \bibinfo {pages} {421} (\bibinfo {year} {2002})}\BibitemShut {NoStop}%
\bibitem [{\citenamefont {Oraevsky}(2002)}]{oraevsky_whispering-gallery_2002}%
  \BibitemOpen
  \bibfield  {author} {\bibinfo {author} {\bibfnamefont {A.~N.}\ \bibnamefont
  {Oraevsky}},\ }\href {\doibase 10.1070/QE2002v032n05ABEH002205} {\bibfield
  {journal} {\bibinfo  {journal} {Quantum Electronics}\ }\textbf {\bibinfo
  {volume} {32}},\ \bibinfo {pages} {377} (\bibinfo {year} {2002})}\BibitemShut
  {NoStop}%
\bibitem [{\citenamefont {Rayleigh}(1879)}]{lord_rayleigh_capillary_1879}%
  \BibitemOpen
  \bibfield  {author} {\bibinfo {author} {\bibfnamefont {L.}~\bibnamefont
  {Rayleigh}},\ }\href@noop {} {\bibfield  {journal} {\bibinfo  {journal}
  {Proc. R. Soc. Lond.}\ }\textbf {\bibinfo {volume} {29}},\ \bibinfo {pages}
  {71} (\bibinfo {year} {1879})}\BibitemShut {NoStop}%
\bibitem [{\citenamefont {Bohr}\ and\ \citenamefont
  {Mottelson}(1998)}]{bohr_nuclear_1998}%
  \BibitemOpen
  \bibfield  {author} {\bibinfo {author} {\bibfnamefont {A.}~\bibnamefont
  {Bohr}}\ and\ \bibinfo {author} {\bibfnamefont {B.~R.}\ \bibnamefont
  {Mottelson}},\ }\href@noop {} {\emph {\bibinfo {title} {Nuclear {Structure}
  {Vol}. {II}: {Nuclear} {Deformations}}}}\ (\bibinfo  {publisher} {World
  Scientific},\ \bibinfo {year} {1998})\BibitemShut {NoStop}%
\bibitem [{\citenamefont {Schmidt}(2015)}]{schmidt_optomechanical_2015}%
  \BibitemOpen
  \bibfield  {author} {\bibinfo {author} {\bibfnamefont {M.}~\bibnamefont
  {Schmidt}},\ }\emph {\bibinfo {title} {Optomechanical arrays and multimode
  systems}},\ \href
  {https://opus4.kobv.de/opus4-fau/frontdoor/index/index/docId/6934} {Ph.D.
  thesis},\ \bibinfo  {school} {University Erlangen-Nuremberg}, \bibinfo
  {address} {Erlangen} (\bibinfo {year} {2015})\BibitemShut {NoStop}%
\bibitem [{\citenamefont {Gang}\ \emph {et~al.}(1995)\citenamefont {Gang},
  \citenamefont {Krall},\ and\ \citenamefont {Weitz}}]{gang_thermal_1995}%
  \BibitemOpen
  \bibfield  {author} {\bibinfo {author} {\bibfnamefont {H.}~\bibnamefont
  {Gang}}, \bibinfo {author} {\bibfnamefont {A.~H.}\ \bibnamefont {Krall}}, \
  and\ \bibinfo {author} {\bibfnamefont {D.~A.}\ \bibnamefont {Weitz}},\ }\href
  {\doibase 10.1103/PhysRevE.52.6289} {\bibfield  {journal} {\bibinfo
  {journal} {Phys. Rev. E}\ }\textbf {\bibinfo {volume} {52}},\ \bibinfo
  {pages} {6289} (\bibinfo {year} {1995})}\BibitemShut {NoStop}%
\bibitem [{\citenamefont {Lai}\ \emph {et~al.}(1990)\citenamefont {Lai},
  \citenamefont {Leung}, \citenamefont {Young}, \citenamefont {Barber},\ and\
  \citenamefont {Hill}}]{lai_time-independent_1990}%
  \BibitemOpen
  \bibfield  {author} {\bibinfo {author} {\bibfnamefont {H.~M.}\ \bibnamefont
  {Lai}}, \bibinfo {author} {\bibfnamefont {P.~T.}\ \bibnamefont {Leung}},
  \bibinfo {author} {\bibfnamefont {K.}~\bibnamefont {Young}}, \bibinfo
  {author} {\bibfnamefont {P.~W.}\ \bibnamefont {Barber}}, \ and\ \bibinfo
  {author} {\bibfnamefont {S.~C.}\ \bibnamefont {Hill}},\ }\href {\doibase
  10.1103/PhysRevA.41.5187} {\bibfield  {journal} {\bibinfo  {journal}
  {Physical Review A}\ }\textbf {\bibinfo {volume} {41}},\ \bibinfo {pages}
  {5187} (\bibinfo {year} {1990})}\BibitemShut {NoStop}%
\bibitem [{\citenamefont {Hill}\ and\ \citenamefont
  {Eaves}(2010)}]{hill_vibrations_2010}%
  \BibitemOpen
  \bibfield  {author} {\bibinfo {author} {\bibfnamefont {R.~J.~A.}\
  \bibnamefont {Hill}}\ and\ \bibinfo {author} {\bibfnamefont {L.}~\bibnamefont
  {Eaves}},\ }\href {\doibase 10.1103/PhysRevE.81.056312} {\bibfield  {journal}
  {\bibinfo  {journal} {Physical Review E}\ }\textbf {\bibinfo {volume} {81}},\
  \bibinfo {pages} {056312} (\bibinfo {year} {2010})}\BibitemShut {NoStop}%
\bibitem [{\citenamefont {Whitaker}\ \emph {et~al.}(1998)\citenamefont
  {Whitaker}, \citenamefont {Kim}, \citenamefont {Vicente}, \citenamefont
  {Weilert}, \citenamefont {Maris},\ and\ \citenamefont
  {Seidel}}]{whitaker_shape_1998}%
  \BibitemOpen
  \bibfield  {author} {\bibinfo {author} {\bibfnamefont {D.~L.}\ \bibnamefont
  {Whitaker}}, \bibinfo {author} {\bibfnamefont {C.}~\bibnamefont {Kim}},
  \bibinfo {author} {\bibfnamefont {C.~L.}\ \bibnamefont {Vicente}}, \bibinfo
  {author} {\bibfnamefont {M.~A.}\ \bibnamefont {Weilert}}, \bibinfo {author}
  {\bibfnamefont {H.~J.}\ \bibnamefont {Maris}}, \ and\ \bibinfo {author}
  {\bibfnamefont {G.~M.}\ \bibnamefont {Seidel}},\ }\href@noop {} {\bibfield
  {journal} {\bibinfo  {journal} {Journal of Low Temperature Physics}\ }\textbf
  {\bibinfo {volume} {113}},\ \bibinfo {pages} {491} (\bibinfo {year}
  {1998})}\BibitemShut {NoStop}%
\bibitem [{\citenamefont {Whitaker}\ \emph {et~al.}(1999)\citenamefont
  {Whitaker}, \citenamefont {Kim}, \citenamefont {Vicente}, \citenamefont
  {Weilert}, \citenamefont {Maris},\ and\ \citenamefont
  {Seidel}}]{whitaker_theory_1999}%
  \BibitemOpen
  \bibfield  {author} {\bibinfo {author} {\bibfnamefont {D.~L.}\ \bibnamefont
  {Whitaker}}, \bibinfo {author} {\bibfnamefont {C.}~\bibnamefont {Kim}},
  \bibinfo {author} {\bibfnamefont {C.~L.}\ \bibnamefont {Vicente}}, \bibinfo
  {author} {\bibfnamefont {M.~A.}\ \bibnamefont {Weilert}}, \bibinfo {author}
  {\bibfnamefont {H.~J.}\ \bibnamefont {Maris}}, \ and\ \bibinfo {author}
  {\bibfnamefont {G.~M.}\ \bibnamefont {Seidel}},\ }\href {\doibase
  10.1023/A:1021810422559} {\bibfield  {journal} {\bibinfo  {journal} {Journal
  of Low Temperature Physics}\ }\textbf {\bibinfo {volume} {114}},\ \bibinfo
  {pages} {523} (\bibinfo {year} {1999})}\BibitemShut {NoStop}%
\bibitem [{\citenamefont {Whitworth}\ \emph {et~al.}(1958)\citenamefont
  {Whitworth}, \citenamefont {Shoneberg},\ and\ \citenamefont
  {S}}]{whitworth_experiments_1958}%
  \BibitemOpen
  \bibfield  {author} {\bibinfo {author} {\bibfnamefont {R.~W.}\ \bibnamefont
  {Whitworth}}, \bibinfo {author} {\bibfnamefont {D.}~\bibnamefont
  {Shoneberg}}, \ and\ \bibinfo {author} {\bibfnamefont {F.~R.}\ \bibnamefont
  {S}},\ }\href {\doibase 10.1098/rspa.1958.0146} {\bibfield  {journal}
  {\bibinfo  {journal} {Proc. R. Soc. Lond. A}\ }\textbf {\bibinfo {volume}
  {246}},\ \bibinfo {pages} {390} (\bibinfo {year} {1958})}\BibitemShut
  {NoStop}%
\bibitem [{\citenamefont {Maris}(1973)}]{maris_hydrodynamics_1973}%
  \BibitemOpen
  \bibfield  {author} {\bibinfo {author} {\bibfnamefont {H.~J.}\ \bibnamefont
  {Maris}},\ }\href {\doibase 10.1103/PhysRevA.8.1980} {\bibfield  {journal}
  {\bibinfo  {journal} {Physical Review A}\ }\textbf {\bibinfo {volume} {8}},\
  \bibinfo {pages} {1980} (\bibinfo {year} {1973})}\BibitemShut {NoStop}%
\bibitem [{\citenamefont {Webeler}\ and\ \citenamefont
  {Hammer}(1965)}]{webeler_viscosity_1965}%
  \BibitemOpen
  \bibfield  {author} {\bibinfo {author} {\bibfnamefont {R.}~\bibnamefont
  {Webeler}}\ and\ \bibinfo {author} {\bibfnamefont {D.}~\bibnamefont
  {Hammer}},\ }\href {\doibase 10.1016/0031-9163(65)90761-4} {\bibfield
  {journal} {\bibinfo  {journal} {Physics Letters}\ }\textbf {\bibinfo {volume}
  {19}},\ \bibinfo {pages} {533} (\bibinfo {year} {1965})}\BibitemShut
  {NoStop}%
\bibitem [{\citenamefont {Webeler}\ and\ \citenamefont
  {Hammer}(1968)}]{webeler_nasa_1968}%
  \BibitemOpen
  \bibfield  {author} {\bibinfo {author} {\bibfnamefont {R.}~\bibnamefont
  {Webeler}}\ and\ \bibinfo {author} {\bibfnamefont {D.}~\bibnamefont
  {Hammer}},\ }\href@noop {} {\emph {\bibinfo {title} {{NASA} {Technical}
  {Note} {D}-4381}}},\ \bibinfo {type} {Tech. Rep.}\ (\bibinfo {year}
  {1968})\BibitemShut {NoStop}%
\bibitem [{\citenamefont {Roche}\ \emph {et~al.}(1996)\citenamefont {Roche},
  \citenamefont {Roger},\ and\ \citenamefont
  {Williams}}]{roche_interpretation_1996}%
  \BibitemOpen
  \bibfield  {author} {\bibinfo {author} {\bibfnamefont {P.}~\bibnamefont
  {Roche}}, \bibinfo {author} {\bibfnamefont {M.}~\bibnamefont {Roger}}, \ and\
  \bibinfo {author} {\bibfnamefont {F.~I.~B.}\ \bibnamefont {Williams}},\
  }\href {\doibase 10.1103/PhysRevB.53.2225} {\bibfield  {journal} {\bibinfo
  {journal} {Phys. Rev. B}\ }\textbf {\bibinfo {volume} {53}},\ \bibinfo
  {pages} {2225} (\bibinfo {year} {1996})}\BibitemShut {NoStop}%
\bibitem [{\citenamefont
  {Chandrasekhar}(1959)}]{chandrasekhar_oscillations_1959}%
  \BibitemOpen
  \bibfield  {author} {\bibinfo {author} {\bibfnamefont {S.}~\bibnamefont
  {Chandrasekhar}},\ }\href {\doibase 10.1112/plms/s3-9.1.141} {\bibfield
  {journal} {\bibinfo  {journal} {Proceedings of the London Mathematical
  Society}\ }\textbf {\bibinfo {volume} {s3-9}},\ \bibinfo {pages} {141}
  (\bibinfo {year} {1959})}\BibitemShut {NoStop}%
\bibitem [{\citenamefont {Betts}\ \emph {et~al.}(1963)\citenamefont {Betts},
  \citenamefont {Osborne}, \citenamefont {Welber},\ and\ \citenamefont
  {Wilks}}]{betts_viscosity_1963}%
  \BibitemOpen
  \bibfield  {author} {\bibinfo {author} {\bibfnamefont {D.~S.}\ \bibnamefont
  {Betts}}, \bibinfo {author} {\bibfnamefont {D.~W.}\ \bibnamefont {Osborne}},
  \bibinfo {author} {\bibfnamefont {B.}~\bibnamefont {Welber}}, \ and\ \bibinfo
  {author} {\bibfnamefont {J.}~\bibnamefont {Wilks}},\ }\href {\doibase
  10.1080/14786436308214457} {\bibfield  {journal} {\bibinfo  {journal}
  {Philosophical Magazine}\ }\textbf {\bibinfo {volume} {8}},\ \bibinfo {pages}
  {977} (\bibinfo {year} {1963})}\BibitemShut {NoStop}%
\bibitem [{\citenamefont {Betts}\ \emph {et~al.}(1965)\citenamefont {Betts},
  \citenamefont {Keen},\ and\ \citenamefont {Wilks}}]{betts_viscosity_1965}%
  \BibitemOpen
  \bibfield  {author} {\bibinfo {author} {\bibfnamefont {D.~S.}\ \bibnamefont
  {Betts}}, \bibinfo {author} {\bibfnamefont {B.~E.}\ \bibnamefont {Keen}}, \
  and\ \bibinfo {author} {\bibfnamefont {J.}~\bibnamefont {Wilks}},\ }\href
  {\doibase 10.1098/rspa.1965.0247} {\bibfield  {journal} {\bibinfo  {journal}
  {Proc. R. Soc. Lond. A}\ }\textbf {\bibinfo {volume} {289}},\ \bibinfo
  {pages} {34} (\bibinfo {year} {1965})}\BibitemShut {NoStop}%
\bibitem [{\citenamefont {Popp}\ \emph {et~al.}(1997)\citenamefont {Popp},
  \citenamefont {Fields},\ and\ \citenamefont {Chang}}]{popp_q_1997}%
  \BibitemOpen
  \bibfield  {author} {\bibinfo {author} {\bibfnamefont {J.}~\bibnamefont
  {Popp}}, \bibinfo {author} {\bibfnamefont {M.~H.}\ \bibnamefont {Fields}}, \
  and\ \bibinfo {author} {\bibfnamefont {R.~K.}\ \bibnamefont {Chang}},\ }\href
  {\doibase 10.1364/OL.22.001296} {\bibfield  {journal} {\bibinfo  {journal}
  {Optics Letters}\ }\textbf {\bibinfo {volume} {22}},\ \bibinfo {pages} {1296}
  (\bibinfo {year} {1997})}\BibitemShut {NoStop}%
\bibitem [{\citenamefont {Lacey}\ and\ \citenamefont
  {Payne}(1990)}]{lacey_radiation_1990}%
  \BibitemOpen
  \bibfield  {author} {\bibinfo {author} {\bibfnamefont {J.~P.~R.}\
  \bibnamefont {Lacey}}\ and\ \bibinfo {author} {\bibfnamefont {F.~P.}\
  \bibnamefont {Payne}},\ }\href@noop {} {\bibfield  {journal} {\bibinfo
  {journal} {IEE Proceedings J - Optoelectronics}\ }\textbf {\bibinfo {volume}
  {137}},\ \bibinfo {pages} {282} (\bibinfo {year} {1990})}\BibitemShut
  {NoStop}%
\bibitem [{\citenamefont {Gorodetsky}\ \emph {et~al.}(2000)\citenamefont
  {Gorodetsky}, \citenamefont {Pryamikov},\ and\ \citenamefont
  {Ilchenko}}]{gorodetsky_rayleigh_2000}%
  \BibitemOpen
  \bibfield  {author} {\bibinfo {author} {\bibfnamefont {M.~L.}\ \bibnamefont
  {Gorodetsky}}, \bibinfo {author} {\bibfnamefont {A.~D.}\ \bibnamefont
  {Pryamikov}}, \ and\ \bibinfo {author} {\bibfnamefont {V.~S.}\ \bibnamefont
  {Ilchenko}},\ }\href {\doibase 10.1364/JOSAB.17.001051} {\bibfield  {journal}
  {\bibinfo  {journal} {JOSA B}\ }\textbf {\bibinfo {volume} {17}},\ \bibinfo
  {pages} {1051} (\bibinfo {year} {2000})}\BibitemShut {NoStop}%
\bibitem [{\citenamefont {Seidel}\ \emph {et~al.}(2002)\citenamefont {Seidel},
  \citenamefont {Lanou},\ and\ \citenamefont {Yao}}]{seidel_rayleigh_2002}%
  \BibitemOpen
  \bibfield  {author} {\bibinfo {author} {\bibfnamefont {G.~M.}\ \bibnamefont
  {Seidel}}, \bibinfo {author} {\bibfnamefont {R.~E.}\ \bibnamefont {Lanou}}, \
  and\ \bibinfo {author} {\bibfnamefont {W.}~\bibnamefont {Yao}},\ }\href
  {\doibase 10.1016/S0168-9002(02)00890-2} {\bibfield  {journal} {\bibinfo
  {journal} {Nuclear Instruments and Methods in Physics Research Section A:
  Accelerators, Spectrometers, Detectors and Associated Equipment}\ }\textbf
  {\bibinfo {volume} {489}},\ \bibinfo {pages} {189} (\bibinfo {year}
  {2002})}\BibitemShut {NoStop}%
\bibitem [{\citenamefont {Weilert}\ \emph {et~al.}(1995)\citenamefont
  {Weilert}, \citenamefont {Whitaker}, \citenamefont {Maris},\ and\
  \citenamefont {Seidel}}]{weilert_laser_1995}%
  \BibitemOpen
  \bibfield  {author} {\bibinfo {author} {\bibfnamefont {M.~A.}\ \bibnamefont
  {Weilert}}, \bibinfo {author} {\bibfnamefont {D.~L.}\ \bibnamefont
  {Whitaker}}, \bibinfo {author} {\bibfnamefont {H.~J.}\ \bibnamefont {Maris}},
  \ and\ \bibinfo {author} {\bibfnamefont {G.~M.}\ \bibnamefont {Seidel}},\
  }\href {\doibase 10.1007/BF00754065} {\bibfield  {journal} {\bibinfo
  {journal} {Journal of Low Temperature Physics}\ }\textbf {\bibinfo {volume}
  {98}},\ \bibinfo {pages} {17} (\bibinfo {year} {1995})}\BibitemShut {NoStop}%
\bibitem [{\citenamefont {Greytak}\ and\ \citenamefont
  {Yan}(1969)}]{greytak_light_1969}%
  \BibitemOpen
  \bibfield  {author} {\bibinfo {author} {\bibfnamefont {T.~J.}\ \bibnamefont
  {Greytak}}\ and\ \bibinfo {author} {\bibfnamefont {J.}~\bibnamefont {Yan}},\
  }\href {\doibase 10.1103/PhysRevLett.22.987} {\bibfield  {journal} {\bibinfo
  {journal} {Physical Review Letters}\ }\textbf {\bibinfo {volume} {22}},\
  \bibinfo {pages} {987} (\bibinfo {year} {1969})}\BibitemShut {NoStop}%
\bibitem [{\citenamefont {Brink}\ and\ \citenamefont
  {Stringari}(1990)}]{brink_density_1990}%
  \BibitemOpen
  \bibfield  {author} {\bibinfo {author} {\bibfnamefont {D.~M.}\ \bibnamefont
  {Brink}}\ and\ \bibinfo {author} {\bibfnamefont {S.}~\bibnamefont
  {Stringari}},\ }\href {\doibase 10.1007/BF01437187} {\bibfield  {journal}
  {\bibinfo  {journal} {Zeitschrift f\{{\textbackslash}"u\}r Physik D Atoms,
  Molecules and Clusters}\ }\textbf {\bibinfo {volume} {15}},\ \bibinfo {pages}
  {257} (\bibinfo {year} {1990})}\BibitemShut {NoStop}%
\bibitem [{Note1()}]{Note1}%
  \BibitemOpen
  \bibinfo {note} {For $^{4}{\protect \rm He}$ vapor pressure at temperatures
  below 0.65K, we use the theoretical Arrhenius law for evaporation rate \cite
  {brink_density_1990}, which fits experimental data well even at more elevated
  temperatures. All other values are found by interpolating published
  experimental data or using published empirical formulas within their region
  of validity.}\BibitemShut {Stop}%
\bibitem [{\citenamefont {Donnelly}\ and\ \citenamefont
  {Barenghi}(1998)}]{donnelly_observed_1998}%
  \BibitemOpen
  \bibfield  {author} {\bibinfo {author} {\bibfnamefont {R.~J.}\ \bibnamefont
  {Donnelly}}\ and\ \bibinfo {author} {\bibfnamefont {C.~F.}\ \bibnamefont
  {Barenghi}},\ }\href {\doibase 10.1063/1.556028} {\bibfield  {journal}
  {\bibinfo  {journal} {Journal of Physical and Chemical Reference Data}\
  }\textbf {\bibinfo {volume} {27}},\ \bibinfo {pages} {1217} (\bibinfo {year}
  {1998})}\BibitemShut {NoStop}%
\bibitem [{\citenamefont {Greywall}(1983)}]{greywall_specific_1983}%
  \BibitemOpen
  \bibfield  {author} {\bibinfo {author} {\bibfnamefont {D.~S.}\ \bibnamefont
  {Greywall}},\ }\href {\doibase 10.1103/PhysRevB.27.2747} {\bibfield
  {journal} {\bibinfo  {journal} {Physical Review B}\ }\textbf {\bibinfo
  {volume} {27}},\ \bibinfo {pages} {2747} (\bibinfo {year}
  {1983})}\BibitemShut {NoStop}%
\bibitem [{\citenamefont {Kerr}(1954)}]{kerr_orthobaric_1954}%
  \BibitemOpen
  \bibfield  {author} {\bibinfo {author} {\bibfnamefont {E.~C.}\ \bibnamefont
  {Kerr}},\ }\href {\doibase 10.1103/PhysRev.96.551} {\bibfield  {journal}
  {\bibinfo  {journal} {Physical Review}\ }\textbf {\bibinfo {volume} {96}},\
  \bibinfo {pages} {551} (\bibinfo {year} {1954})}\BibitemShut {NoStop}%
\bibitem [{\citenamefont {Huang}\ and\ \citenamefont
  {Chen}(2006)}]{huang_practical_2006}%
  \BibitemOpen
  \bibfield  {author} {\bibinfo {author} {\bibfnamefont {Y.~H.}\ \bibnamefont
  {Huang}}\ and\ \bibinfo {author} {\bibfnamefont {G.~B.}\ \bibnamefont
  {Chen}},\ }\href {\doibase 10.1016/j.cryogenics.2006.07.006} {\bibfield
  {journal} {\bibinfo  {journal} {Cryogenics}\ }\textbf {\bibinfo {volume}
  {46}},\ \bibinfo {pages} {833} (\bibinfo {year} {2006})}\BibitemShut
  {NoStop}%
\bibitem [{\citenamefont {Clerk}\ \emph {et~al.}(2010)\citenamefont {Clerk},
  \citenamefont {Devoret}, \citenamefont {Girvin}, \citenamefont {Marquardt},\
  and\ \citenamefont {Schoelkopf}}]{clerk_introduction_2010}%
  \BibitemOpen
  \bibfield  {author} {\bibinfo {author} {\bibfnamefont {A.~A.}\ \bibnamefont
  {Clerk}}, \bibinfo {author} {\bibfnamefont {M.~H.}\ \bibnamefont {Devoret}},
  \bibinfo {author} {\bibfnamefont {S.~M.}\ \bibnamefont {Girvin}}, \bibinfo
  {author} {\bibfnamefont {F.}~\bibnamefont {Marquardt}}, \ and\ \bibinfo
  {author} {\bibfnamefont {R.~J.}\ \bibnamefont {Schoelkopf}},\ }\href
  {\doibase 10.1103/RevModPhys.82.1155} {\bibfield  {journal} {\bibinfo
  {journal} {Reviews of Modern Physics}\ }\textbf {\bibinfo {volume} {82}},\
  \bibinfo {pages} {1155} (\bibinfo {year} {2010})}\BibitemShut {NoStop}%
\bibitem [{\citenamefont {Busse}(1984)}]{busse_oscillations_1984}%
  \BibitemOpen
  \bibfield  {author} {\bibinfo {author} {\bibfnamefont {F.~H.}\ \bibnamefont
  {Busse}},\ }\href {\doibase 10.1017/S0022112084000963} {\bibfield  {journal}
  {\bibinfo  {journal} {Journal of Fluid Mechanics}\ }\textbf {\bibinfo
  {volume} {142}},\ \bibinfo {pages} {1} (\bibinfo {year} {1984})}\BibitemShut
  {NoStop}%
\bibitem [{\citenamefont {Bryan}(1890)}]{bryan_onthe_1890}%
  \BibitemOpen
  \bibfield  {author} {\bibinfo {author} {\bibfnamefont {G.~H.}\ \bibnamefont
  {Bryan}},\ }\href@noop {} {\bibfield  {journal} {\bibinfo  {journal} {Proc.
  Cambridge Phil. Soc.}\ }\textbf {\bibinfo {volume} {VII}},\ \bibinfo {pages}
  {101} (\bibinfo {year} {1890})}\BibitemShut {NoStop}%
\bibitem [{\citenamefont {Joubert}\ \emph {et~al.}(2009)\citenamefont
  {Joubert}, \citenamefont {Shatalov},\ and\ \citenamefont
  {Fay}}]{joubert_rotating_2009}%
  \BibitemOpen
  \bibfield  {author} {\bibinfo {author} {\bibfnamefont {S.~V.}\ \bibnamefont
  {Joubert}}, \bibinfo {author} {\bibfnamefont {M.~Y.}\ \bibnamefont
  {Shatalov}}, \ and\ \bibinfo {author} {\bibfnamefont {T.~H.}\ \bibnamefont
  {Fay}},\ }\href {\doibase 10.1119/1.3088877} {\bibfield  {journal} {\bibinfo
  {journal} {American Journal of Physics}\ }\textbf {\bibinfo {volume} {77}},\
  \bibinfo {pages} {520} (\bibinfo {year} {2009})}\BibitemShut {NoStop}%
\bibitem [{\citenamefont {Greenspan}(1968)}]{greenspan_theory_1968}%
  \BibitemOpen
  \bibfield  {author} {\bibinfo {author} {\bibfnamefont {H.~P.}\ \bibnamefont
  {Greenspan}},\ }\href@noop {} {\emph {\bibinfo {title} {The {Theory} of
  {Rotating} {Fluids}}}}\ (\bibinfo  {publisher} {CUP Archive},\ \bibinfo
  {year} {1968})\ \bibinfo {note} {google-Books-ID: 2R47AAAAIAAJ}\BibitemShut
  {NoStop}%
\bibitem [{\citenamefont {Landau}\ and\ \citenamefont
  {Lifshitz}(2013)}]{landau_fluid_2013}%
  \BibitemOpen
  \bibfield  {author} {\bibinfo {author} {\bibfnamefont {L.~D.}\ \bibnamefont
  {Landau}}\ and\ \bibinfo {author} {\bibfnamefont {E.~M.}\ \bibnamefont
  {Lifshitz}},\ }\href@noop {} {\emph {\bibinfo {title} {Fluid {Mechanics}}}}\
  (\bibinfo  {publisher} {Elsevier},\ \bibinfo {year} {2013})\ \bibinfo {note}
  {google-Books-ID: CeBbAwAAQBAJ}\BibitemShut {NoStop}%
\bibitem [{\citenamefont {Bahl}\ \emph {et~al.}(2012)\citenamefont {Bahl},
  \citenamefont {Tomes}, \citenamefont {Marquardt},\ and\ \citenamefont
  {Carmon}}]{bahl_observation_2012}%
  \BibitemOpen
  \bibfield  {author} {\bibinfo {author} {\bibfnamefont {G.}~\bibnamefont
  {Bahl}}, \bibinfo {author} {\bibfnamefont {M.}~\bibnamefont {Tomes}},
  \bibinfo {author} {\bibfnamefont {F.}~\bibnamefont {Marquardt}}, \ and\
  \bibinfo {author} {\bibfnamefont {T.}~\bibnamefont {Carmon}},\ }\href
  {\doibase 10.1038/nphys2206} {\bibfield  {journal} {\bibinfo  {journal}
  {Nature Physics}\ }\textbf {\bibinfo {volume} {8}},\ \bibinfo {pages} {203}
  (\bibinfo {year} {2012})}\BibitemShut {NoStop}%
\bibitem [{\citenamefont {Hill}\ and\ \citenamefont
  {Eaves}(2008)}]{hill_nonaxisymmetric_2008}%
  \BibitemOpen
  \bibfield  {author} {\bibinfo {author} {\bibfnamefont {R.~J.~A.}\
  \bibnamefont {Hill}}\ and\ \bibinfo {author} {\bibfnamefont {L.}~\bibnamefont
  {Eaves}},\ }\href {\doibase 10.1103/PhysRevLett.101.234501} {\bibfield
  {journal} {\bibinfo  {journal} {Physical Review Letters}\ }\textbf {\bibinfo
  {volume} {101}},\ \bibinfo {pages} {234501} (\bibinfo {year}
  {2008})}\BibitemShut {NoStop}%
\bibitem [{\citenamefont {Heine}(2006)}]{heine_computations_2006}%
  \BibitemOpen
  \bibfield  {author} {\bibinfo {author} {\bibfnamefont {C.-J.}\ \bibnamefont
  {Heine}},\ }\href {\doibase 10.1093/imanum/drl007} {\bibfield  {journal}
  {\bibinfo  {journal} {IMA Journal of Numerical Analysis}\ }\textbf {\bibinfo
  {volume} {26}},\ \bibinfo {pages} {723} (\bibinfo {year} {2006})}\BibitemShut
  {NoStop}%
\bibitem [{\citenamefont {Seidel}\ and\ \citenamefont
  {Maris}(1994)}]{seidel_morphology_1994}%
  \BibitemOpen
  \bibfield  {author} {\bibinfo {author} {\bibfnamefont {G.~M.}\ \bibnamefont
  {Seidel}}\ and\ \bibinfo {author} {\bibfnamefont {H.~J.}\ \bibnamefont
  {Maris}},\ }\href@noop {} {\bibfield  {journal} {\bibinfo  {journal} {Physica
  B}\ }\textbf {\bibinfo {volume} {194-196}},\ \bibinfo {pages} {577} (\bibinfo
  {year} {1994})}\BibitemShut {NoStop}%
\bibitem [{\citenamefont {Vinen}\ \emph {et~al.}(2003)\citenamefont {Vinen},
  \citenamefont {Tsubota},\ and\ \citenamefont
  {Mitani}}]{vinen_kelvin-wave_2003}%
  \BibitemOpen
  \bibfield  {author} {\bibinfo {author} {\bibfnamefont {W.~F.}\ \bibnamefont
  {Vinen}}, \bibinfo {author} {\bibfnamefont {M.}~\bibnamefont {Tsubota}}, \
  and\ \bibinfo {author} {\bibfnamefont {A.}~\bibnamefont {Mitani}},\
  }\href@noop {} {\bibfield  {journal} {\bibinfo  {journal} {Physical Review
  Letters}\ }\textbf {\bibinfo {volume} {91}},\ \bibinfo {pages} {135301}
  (\bibinfo {year} {2003})}\BibitemShut {NoStop}%
\bibitem [{\citenamefont {Kozik}\ and\ \citenamefont
  {Svistuniv}(2004)}]{kozik_kelvin-wave_2004}%
  \BibitemOpen
  \bibfield  {author} {\bibinfo {author} {\bibfnamefont {E.}~\bibnamefont
  {Kozik}}\ and\ \bibinfo {author} {\bibfnamefont {B.}~\bibnamefont
  {Svistuniv}},\ }\href@noop {} {\bibfield  {journal} {\bibinfo  {journal}
  {Physical Review Letters}\ }\textbf {\bibinfo {volume} {92}},\ \bibinfo
  {pages} {035301} (\bibinfo {year} {2004})}\BibitemShut {NoStop}%
\bibitem [{\citenamefont {Volovik}(2009)}]{volovik_universe_2009}%
  \BibitemOpen
  \bibfield  {author} {\bibinfo {author} {\bibfnamefont {G.~E.}\ \bibnamefont
  {Volovik}},\ }\href@noop {} {\emph {\bibinfo {title} {The {Universe} in a
  {Helium} {Droplet}}}},\ \bibinfo {edition} {1st}\ ed.\ (\bibinfo  {publisher}
  {Oxford University Press},\ \bibinfo {address} {Oxford},\ \bibinfo {year}
  {2009})\BibitemShut {NoStop}%
\end{thebibliography}%

\end{document}